\documentclass[aps,prb,reprint,superscriptaddress]{revtex4-1}

\usepackage{amsmath}
\usepackage{graphicx}
\usepackage{xcolor}
\usepackage[normalem]{ulem}

\DeclareMathOperator{\Tr}{Tr}
\newcommand{\udket}{{|\! \uparrow \downarrow \rangle}}
\newcommand{\duket}{{|\! \downarrow \uparrow \rangle}}
\newcommand{\uuket}{{|\! \uparrow \uparrow \rangle}}
\newcommand{\ddket}{{|\! \downarrow \downarrow \rangle}}

\newcommand{\swap}{\textsc{swap}}
\newcommand{\cz}{\textsc{cz}}
\begin{document}

%\preprint{}

\title{Protecting Quantum Information in Quantum Dot Spin Chains by Driving Exchange Interactions Periodically}

\author{John S. Van Dyke}
\affiliation{Department of Physics, Virginia Tech, Blacksburg, Virginia 24061, USA}
\author{Yadav  P.  Kandel}
\affiliation{Department of Physics and Astronomy, University of Rochester, Rochester, NY, 14627 USA}
\author{Haifeng  Qiao}
\affiliation{Department of Physics and Astronomy, University of Rochester, Rochester, NY, 14627 USA}
\author{John M. Nichol}
\affiliation{Department of Physics and Astronomy, University of Rochester, Rochester, NY, 14627 USA}
\author{Sophia E. Economou}
\affiliation{Department of Physics, Virginia Tech, Blacksburg, Virginia 24061, USA}

\author{Edwin Barnes}
\affiliation{Department of Physics, Virginia Tech, Blacksburg, Virginia 24061, USA}

\date{\today}

\begin{abstract}
Recent work has demonstrated a new route to discrete time crystal physics in quantum spin chains by periodically driving nearest-neighbor exchange interactions in gate-defined quantum dot arrays [arXiv:2006.10913]. Here, we present a detailed analysis of exchange-driven Floquet physics in small arrays of GaAs quantum dots, including phase diagrams and additional diagnostics. We also show that emergent time-crystalline behavior can benefit the protection and manipulation of multi-spin states. For typical levels of nuclear spin noise in GaAs, the combination of driving and interactions protects spin-singlet states beyond what is possible in the absence of exchange interactions. We further show how to construct a time-crystal-inspired \cz\ gate between singlet-triplet qubits with high fidelity. These results show that periodically driving exchange couplings can enhance the performance of quantum dot spin systems for quantum information applications.
\end{abstract}

%\keywords{}

\maketitle


\section{Introduction}
Rapid theoretical and experimental development of quantum computers has led to a productive crossover of ideas between the fields of many-body condensed matter physics and of quantum information and computation \cite{Augusiak2012,Zeng2019}. On the one hand, a principal application of quantum devices is the simulation of quantum many-body systems that are not amenable to classical computational methods \cite{Preskill2018,McClean2016,Kandala2017}.   However, the relationship is not merely one-way: concepts from many-body physics can also be useful in designing new quantum devices with improved information processing capabilities.  This direction is exemplified by recent work on many-body localization, time crystals, and fractons \cite{Else2016,Yao2017,Abanin2019,Else2020,Khemani2020,Khemani2019}, which have been variously proposed for robust storage of quantum information \cite{Yao2015,Santos2020}. 

Studies of discrete time crystals (DTCs) in spin systems have largely employed single-spin rotations as the driving terms that are needed to realize the DTC phase \cite{Else2016,Yao2017,Zhang2017,Choi2017}.  Such driving can be achieved in quantum dots (QDs), for instance, by electric dipole spin resonance (EDSR) via an embedded micromagnet \cite{PioroLadriere2008,Watson2018,Sigillito2019,Takeda2020}.  But gate-defined QDs also afford exquisite control over spin interactions, whether by detuning or symmetric barrier gates \cite{Petta2005,Reed2016,Martins2016}.  This motivates the exploration of novel driving protocols in which the spin interactions are periodically modulated. Driving the interactions also allows one to implement important  operations, such as a \swap\ between the states of neighboring QD spins, which is useful for measuring states in the middle of an array by shuttling the desired state to the edge for readout. A recent paper has developed a \swap\ DTC driving protocol in which exchange driving of spin pairs by \swap\ operations, followed by periods of weak interaction, produces time-crystal-like signatures in a four spin QD array \cite{Qiao2020}.  

In this paper, we explore the preservation and manipulation of entanglement in QD spin chains via the \swap\ DTC protocol.  We show that arbitrary states in the $S_z=0$ subspace of two neighboring spins can be preserved for long times, with marked improvement over the undriven interacting system. This result, obtained for finite chains, is reminiscent of DTC physics in the thermodynamic limit, due to the crucial role played by interactions in stabilizing the state.  It also suggests the application of the \swap\ DTC protocol as a form of dynamic quantum memory, protecting the state of the two entangled spins.  One may further consider such pairs of neighboring spins as forming singlet-triplet (ST) qubits \cite{Levy2002,Petta2005}.  For this case, we design a universal gate set, which includes a high-fidelity \cz\ gate through the modification of the \swap\ DTC protocol. Taken together, these results show that DTC-based physics offers a promising route for developing quantum information processing systems in solid-state spin arrays.

The paper is structured as follows.  Section~\ref{sec:model} introduces the model and the driving protocol for the \swap\ DTC.  Section~\ref{sec:PDs} presents phase diagrams that demonstrate the robustness of the DTC phase to the presence of driving errors, a key requirement for the \swap\ DTC to constitute a genuine phase of matter and to be of practical use.  In Section~\ref{sec:retprob}, we investigate the time dependence of the return probability and uncover the existence of $4T$ periodic oscillations for initial entangled spin states, in contrast with the usual $2T$ time translation symmetry breaking found in earlier studies.  Section~\ref{sec:undriven} compares the return probabilities for different driving protocols and for the undriven Heisenberg spin chain, illustrating the importance of driving for preserving entangled states of the two spins in an ST qubit.  Section~\ref{sec:switchstate} demonstrates the single-qubit gate allowing for coherent switching of the preserved state.  Section~\ref{sec:twoqubitgates} describes the \cz\ gate inspired by the \swap\ DTC protocol and presents numerical calculations of its fidelity.  Finally, the results are summarized in Section~\ref{sec:conclusion}.

\section{Model of a \swap\ Time Crystal \label{sec:model}}

We consider a one-dimensional chain of spin-1/2 degrees of freedom consisting of $L = 2N_q$ sites. The Hamiltonian for this system is given by
\begin{align}
H = \sum_{\langle ij \rangle,\alpha} \frac{J_{ij}}{4} \sigma^\alpha_i \sigma^\alpha_j + \sum_i \frac{1}{2}(B_0 + \delta B_i ) \sigma^z_i, \label{eq:heisham}
\end{align}
where $\alpha = \{x,y,z\}$ and $\langle ij \rangle$ indicates nearest-neighbors.  $J_{ij}$ is the exchange interaction, $B_0$ is an externally applied uniform magnetic field, and $\delta B_i$ is a random Gaussian-distributed contribution to the total field with variance $\sigma_B$ due to nuclear spin noise (as in GaAs, for instance). 

\begin{figure}[h]
\includegraphics[scale=0.8]{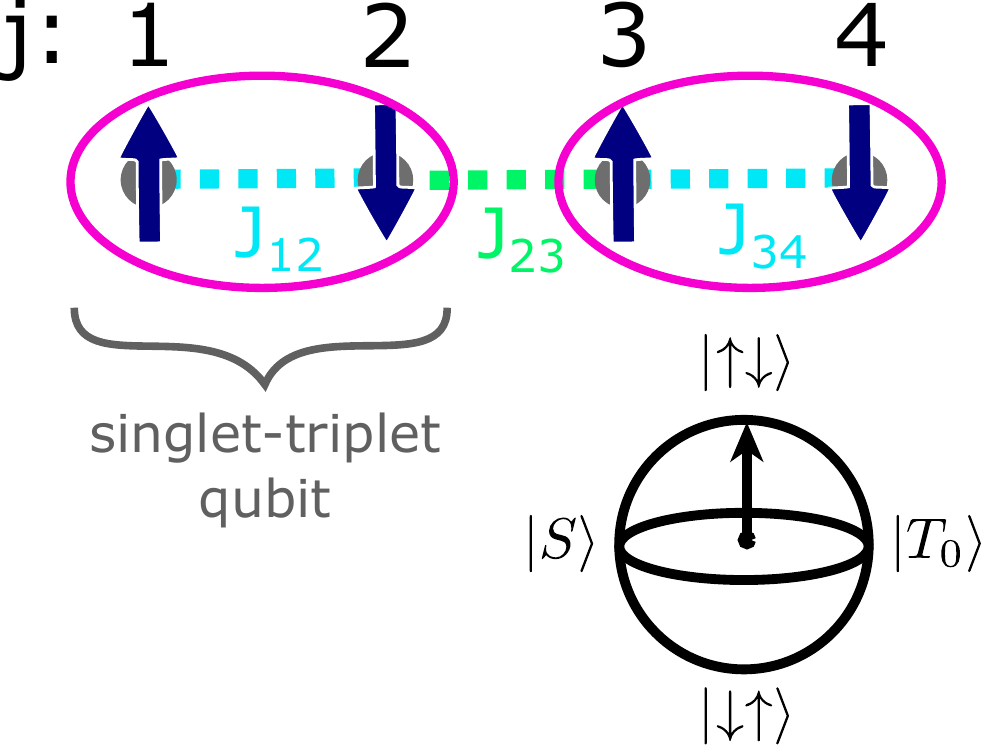}
\caption{ Schematic of an $L=4$ Heisenberg spin chain with variable exchange interactions $J_{12}$, $J_{23}$, and $J_{34}$.  One can think of this system as a pair of coupled ST qubits (with leakage), as indicated by the purple ovals.  $J_{12}$ and $J_{34}$ are used to execute \swap\ operations on the spins defining these qubits, while $J_{23}$ yields an interaction between them.  The ST qubit Bloch sphere is also shown. \label{fig:schematic}}
\end{figure}

Although the principles we discuss apply to generic spin-1/2 Heisenberg chains, we find it helpful to think of the system as an array of coupled ST qubits \cite{Levy2002}. An ST qubit consists of a pair of electron spins on neighboring QDs subject to a large magnetic field that separates out the polarized states, $|T_+ \rangle = | \! \!\uparrow \uparrow \rangle$ and $|T_- \rangle = | \! \!\downarrow \downarrow \rangle$, leaving behind the computational subspace $\{ |S \rangle, | T_0 \rangle \}$ of the singlet ($|S \rangle = (| \! \!\uparrow \downarrow \rangle - |\! \!\downarrow \uparrow \rangle )/\sqrt{2}$)) and $S_z = 0$ triplet ($|T_0 \rangle = (|\! \!\uparrow \downarrow \rangle + |\! \!\downarrow \uparrow \rangle )/\sqrt{2}$)) states.  The resulting two-level system admits a Bloch sphere representation, as shown in Fig. \ref{fig:schematic}, where the basis $\{ \udket, \duket \}$ is chosen for the $\hat{z}$ direction. ST qubits are actively being studied as an encoding for qubits that are naturally insensitive to uniform magnetic field fluctuations \cite{Petta2005,Shulman2012,Wang2012,Calderon-Vargas2015,Nichol2017,Buterakos2019,Colmenar2019,Cerfontaine2020a}. $N_q$ is the number of ST qubits in the chain, which are comprised of pairs of neighboring sites $(2q-1,2q)$, with $q=1,2,...$ (Fig. \ref{fig:schematic}).  

Time crystalline phases were previously discovered in driven Heisenberg chains by applying tailored ``H2I'' pulse sequences or magnetic field gradients that convert the Heisenberg interactions into effective Ising ones \cite{Barnes2019,Li2020}. In both approaches, the periodic driving consisted of single-particle terms that rotate the spins by $\pi$, whether by idealized $\delta$-function pulses or realistic EDSR methods.  Notably, it was found to be necessary to apply H2I pulses or field gradients in order to stabilize a DTC for the levels of magnetic field noise present in experiment (e.g. 18 MHz in GaAs, such that $T_2^* \approx 10$ ns).  

Here, we consider a driving protocol based on varying the exchange interactions in a QD array, instead of single-spin manipulations.  This approach has several advantages.  For one, it can be performed in systems that lack the micromagnet needed for EDSR.  More importantly, the timescales for modifying the nearest-neighbor exchange are very fast (a few nanoseconds), whereas EDSR is slower for the weak to moderate field gradients typically used in experiment \cite{PioroLadriere2008}.  The fundamental idea of our approach is to drive the system periodically by fast \swap\ operations within each ST qubit, followed by long evolution times during which neighboring ST qubits interact \cite{Qiao2020}.  Both of these operations are implemented by the same underlying physical mechanism, namely, the nearest-neighbor exchange coupling between QD spins. More specifically, we consider the following unitary evolution over one drive period:
\begin{align}
U = U_{SWAP}(T_{S}) U_{evo}(T_{e}) . \label{eq:swapDTCprotocol}
\end{align}
The two parts of this protocol are piecewise constant, with the \swap\ piece given by $U_{SWAP}(T_{S}) = e^{-i H_S T_S}$, where
\begin{align}
H_S = \frac{J_{S}}{4}(1-\epsilon) \sum_{i=1,\alpha}^{L/2}  \sigma^\alpha_{2i-1} \sigma^\alpha_{2i} + \sum_{i=1}^L \frac{1}{2}(B_0 + \delta B_i ) \sigma^z_i
\end{align}

is applied for time $T_S$ such that $J_S T_S = \pi$, thus interchanging the spin states of sites $2i-1$ and $2i$.  $\epsilon$ introduces a fractional error in the \swap\ pulse, corresponding to an underrotation for $\epsilon > 0$. For the $L=4$ chain, the \swap\ interactions are illustrated by the light blue dashed lines in Fig. \ref{fig:schematic}, such that $J_{12} = J_{34} = J_S$. The evolution piece $U_{evo}(T_{e}) = e^{-i H_e T_e}$ is generated by the Hamiltonian
\begin{align}
H_e= \frac{J_{e}}{4} \sum_{i=1,\alpha}^{L/2-1}  \sigma^\alpha_{2i} \sigma^\alpha_{2i+1} + \sum_{i=1}^L \frac{1}{2}(B_0 + \delta B_i ) \sigma^z_i. \label{eq:Hevo}
\end{align}
These interactions are indicated by the light green dashed line in Fig. \ref{fig:schematic}, with $J_{23} = J_e$.  In the following sections, we explore the consequences of this driving protocol for the stabilization of quantum information.  Unless otherwise stated, we assume an $L=4$ chain in our numerical calculations.  The calculations were performed using the QuSpin Python package for exact diagonalization of quantum many-body systems \cite{Weinberg2017}.

\section{Phase Diagrams \label{sec:PDs}}
\begin{figure}[h]
\includegraphics[scale=0.8]{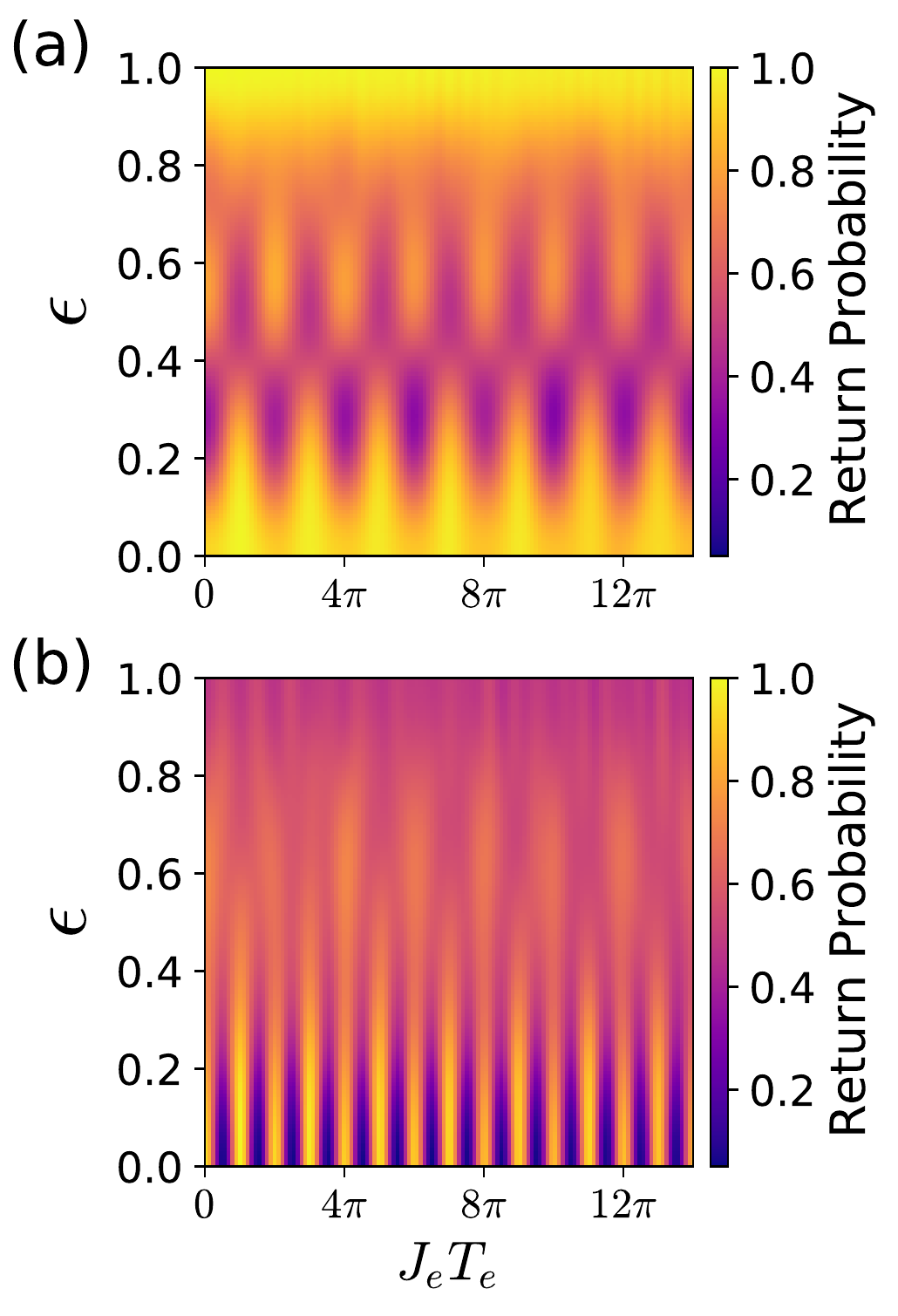}
\caption{ (a) Phase diagram of the return probability for an initial $\udket$ state on qubit 1 as a function of inter-qubit coupling $J_e$ and pulse error $\epsilon$.  (b) Phase diagram of the return probability for an initial singlet state of qubit 1.  Parameters are $L=4$, $B_0 = 3075$ MHz, $\sigma_B = 18$ MHz, $T_{e}=1.4$ $\mu$s, $T_S=2$ ns, $J_S = \pi/T_S$, $n_T$=4.  Here we have chosen parameters similar to those of Ref. \cite{Qiao2020}.  The initial state of qubit 2 is the product state that minimizes the field gradient energy for a given disorder realization.  Results are averaged over 192 disorder realizations. \label{fig:PD}}
\end{figure}

One of the defining features of a time crystal is its stability to perturbations due to the presence of non-zero interactions in the system.  Earlier work on both Ising model and Heisenberg model DTCs has shown that sufficiently weak driving pulse errors (i.e. over- or under-rotation of the spins relative to $\pi$ radians) do not destroy the phase.  Here we examine the corresponding errors in performing an incomplete \swap\ operation.  Fig.~\ref{fig:PD}(a) shows the subsystem return probability for qubit 1 (sites 1 and 2) of an $L=4$ spin chain, after four periods of the protocol ($n_T=4$).  The system is initialized in the product state in which each ST qubit is in its individual non-interacting ground state, the latter being determined by the local magnetic field gradient across the double QD.  Thus, the initial state chosen varies over the field noise disorder realizations.  This scenario is naturally realized in experiments with gate-defined QD arrays.  In our calculations, we fix the evolution time to $T_e = 1.4$ $\mu$s, and we vary the interaction strength $J_e$ and the fractional error in performing a \swap, i.e. an error of $\epsilon = 0.5$ corresponds to a $\sqrt{\swap}$, while for $\epsilon=1$ no operation is performed at all. We find that typical levels of charge noise have little effect on the results, so we neglect this here.
The wedge-shaped regions of high return probability for small $\epsilon$ and increasing $J_e$ illustrate that interactions are crucial for preserving the quantum state of qubit 1 in the presence of driving errors.  We note that not driving the system at all ($\epsilon = 1$) is also very effective for preserving the state of qubit 1 (though of course in this case there is no time translation symmetry breaking).  We examine this further in Section~\ref{sec:undriven}.  

\begin{figure}[h]
\includegraphics[scale=0.4]{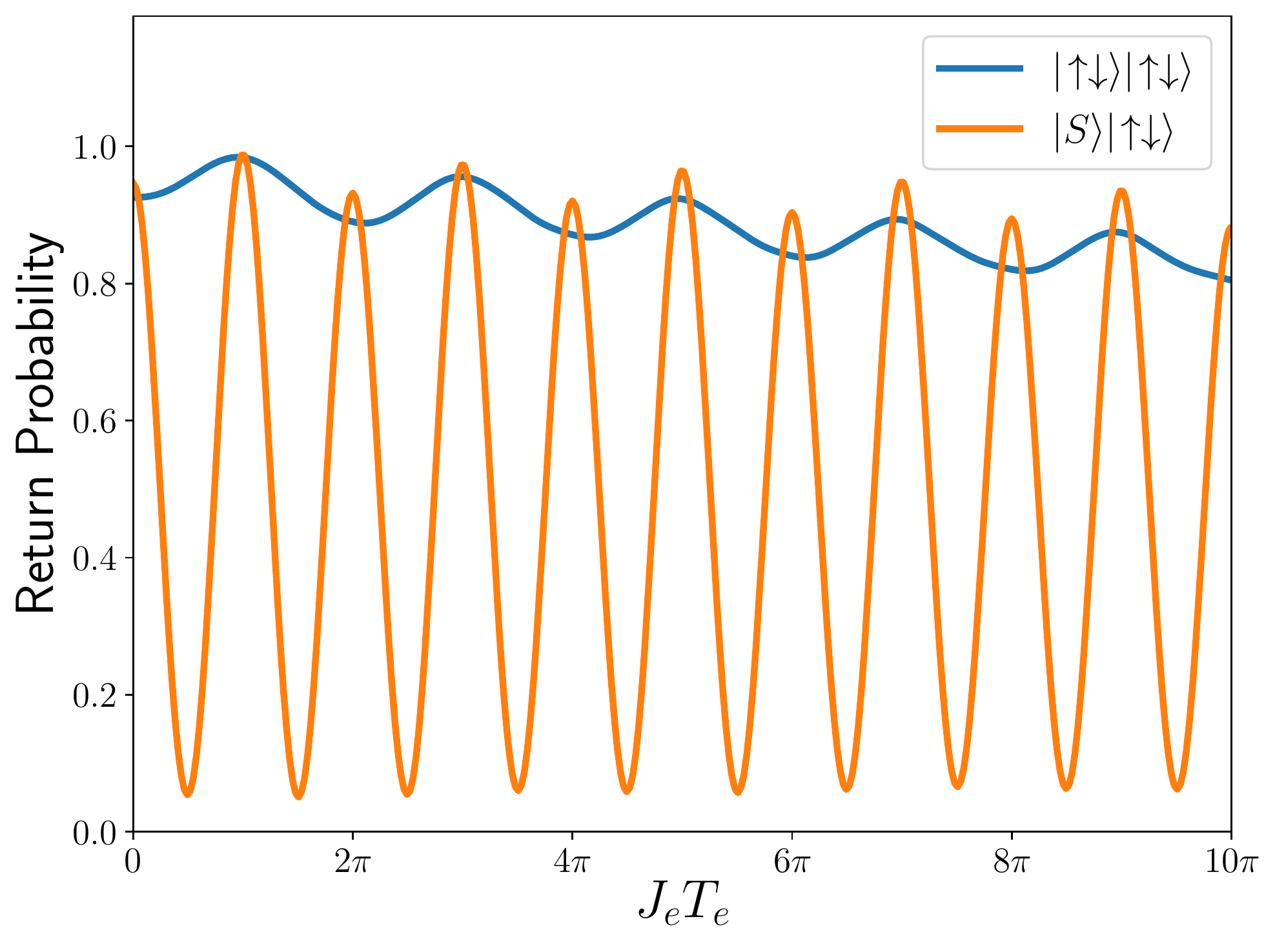}
\caption{ Return probability for qubit 1 as a function of $J_e T_e$, for the initial states $\udket$ (blue line) and the singlet (orange line). Parameters are $L=4$, $B_0 = 3075$ MHz, $\sigma_B = 18$ MHz, $J_S = 250$ MHz, $T_{e}=1.4$ $\mu$s, $T_S=2$ ns, $n_T$=4, $\epsilon = 0$. The initial state of qubit 2 is $\udket$. Results are averaged over 960 disorder realizations. \label{fig:PD1D}}
\end{figure}

In contrast, Fig. \ref{fig:PD}(b) reveals that when qubit 1 is initialized in a singlet state, \swap\ driving is required to produce a high return probability after four periods of evolution.  Here, the initial state of qubit 2 is still the product state determined by the local field gradient. While $J_e=0$ yields a high singlet return probability for a perfect \swap, the presence of finite interactions does increase the value of the return probability, as seen in Fig. \ref{fig:PD1D}.  The singlet return probability peaks when $J_e T_e = \pi n$ (for $J_e$ measured in rad/$\mu$s).  In weak magnetic field gradients, these values correspond to performing $n$ \swap\ operations on sites belonging to different neighboring qubits (e.g. sites 2 and 3 in the $L=4$ chain).  An even $n$ yields a net trivial operation (for perfect \swap s), while odd $n$ causes the initial singlet on sites 1 and 2, $S_{12}$, to be transferred to sites 1 and 3 during the evolution piece of the protocol, which is then undone after three additional periods in the $L=4$ case.  The low values of $S_{12}$ in between the peaks can be understood as arising from the monogamy of entanglement, since an incomplete \swap\ leads to site 1 remaining partially entangled with the rest of the chain after four periods, and thus less entangled with site 2.  When the initial state is the product state $\udket \udket$, the \swap\ on 2 and 3 produces a spin echo-like effect that accounts for the maxima when $n$ is odd.

\section{Return Probability Dynamics \label{sec:retprob}}

\begin{figure}[h]
\includegraphics[scale=0.89]{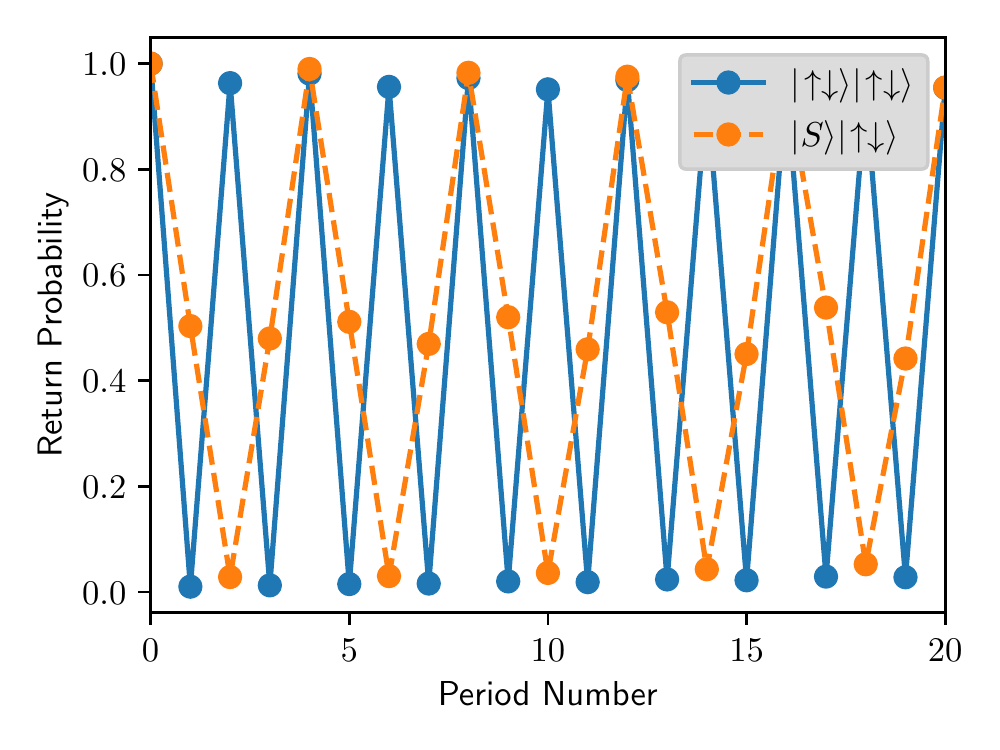}
\caption{ Time dependence of the return probabilities for qubit 1, given an initial state of $\udket$ for qubit 2.  The blue line shows the return probability for $\udket$, given the initial product state $\udket \udket$.  The orange shows the return probability for the singlet state $|S \rangle$, given the initial product state $|S \rangle \udket$ (orange line).  Parameters are $L=4$, $B_0 = 3075$ MHz, $\sigma_B = 18$ MHz, $T_e = 1.4$ $\mu$s, $J_e=\pi/T_e$, $T_S=2$ ns, $J_S=\pi/T_S$, $\epsilon = 0$.  Results are averaged over 6000 disorder realizations.  \label{fig:retprobvst_udS}}
\end{figure}

The dynamics are also different depending on whether the initial state is a product or singlet state.  Fig. \ref{fig:retprobvst_udS} illustrates the $2T$ periodicity of  the return probability for qubit 1 when the system is initialized in $\udket \udket$ and $J_e T_e = \pi$.  The results agree with those for a chain driven by single-spin $\pi$ rotations, as both operations have the same effect: $\udket \udket \rightarrow \duket \duket$.

On the other hand, the $L=4$ chain shows a $4T$ periodicity for the singlet return probability of qubit 1.  This is in striking contrast with previous work on discrete time crystals, which generally found a $2T$ periodicity for spin-1/2 degrees of freedom \cite{Else2016,Yao2017,Zhang2017,Choi2017}.  In fact, for $\sigma_B \ll J_e$ we find that an $L$ site chain has a singlet return probability with $LT$ periodicity.  This can be easily understood as arising from successive applications of \swap s, coming from both the explicit driving part of the protocol and the evolution part tuned to $J_e T_e = \pi$.  For instance, when $L=6$ we have the following steps that transfer the singlet state down the chain, where it is ``reflected'' off the right edge and returns back to its initial position: 
\begin{align}
S_{12} &\xrightarrow{\swap} S_{12} \xrightarrow{evo} S_{13} \xrightarrow{\swap} S_{24} \xrightarrow{evo} S_{35} \notag\\
&\xrightarrow{\swap} S_{46} \xrightarrow{evo} S_{56} \xrightarrow{\swap} S_{56} \xrightarrow{evo} S_{46}\notag\\
&\xrightarrow{\swap} S_{35} \xrightarrow{evo} S_{24} \xrightarrow{\swap} S_{13} \xrightarrow{evo} S_{12}
\end{align}

However, the experimentally relevant interaction strength needed to perform a single \swap\ over $1.4$ $\mu$s is $\sim 350$ kHz, which is much smaller than the magnetic field noise $\sim 18$ MHz in GaAs QDs.  For realistic levels of field noise, the singlet return probability displays a $4T$ periodicity regardless of chain length. Moreover, we find that when the disorder starts at small values and increases toward 18 MHz, the transition between $6T$ and $4T$ periodicity is smooth, with the return probability at $6T$ gradually decreasing while that at $4T$ increases (as opposed to a shift in the peak from $6T$ to $4T$ through intermediate values).  

The $4T$ periodicity observed at sufficiently strong disorder can be explained as follows. First, note that each part of the protocol involves interactions only between disjoint pairs of spins.  Thus, we may consider the Hamiltonian, Eq.~\eqref{eq:heisham}, restricted to two sites $a$ and $b$,
\begin{align}
H_{ab} = \frac{J}{4} ( \sigma^x_a \sigma^x_b +  \sigma^y_a \sigma^y_b +  \sigma^z_a \sigma^z_b) + \frac{1}{2} ( B_a \sigma^z_a + B_b \sigma^z_b ),
\end{align}
where $B_{a,b}$ is the total field at site $a,b$.  In general, the two spins coupled in a given part of the protocol can have parallel or antiparallel orientations. Within the  $\{ \udket, \duket \}$ subspace the evolution operator $U = e^{-it H_{ab}}$ is
\begin{align}
U_1 = e^{iJt/2} \begin{pmatrix}
  \cos ( \frac{\alpha t}{2} )+  \frac{i \Delta}{\alpha}  \sin(t\alpha/2)  & -\frac{i  J}{\alpha} \sin (\frac{\alpha t}{2}) \\
 -\frac{i  J}{\alpha} \sin (\frac{\alpha t}{2}) &   \cos (\frac{\alpha}{2} ) - \frac{i \Delta}{\alpha}  \sin(\frac{\alpha t}{2})
\end{pmatrix},
\end{align}
with $\alpha = \sqrt{J^2 + \Delta^2}$ and $\Delta = B_b - B_a$ the field gradient across the pair. We have multiplied $U$ (and hence $U_1$) by a global phase, $e^{iJt/4}$, to simplify the following analysis.  The \swap\ part of the protocol is performed in 2 ns, so that $J_S \gg \Delta$ and we may neglect errors in the transition $\udket \xrightarrow{\swap} \duket$.  For the evolution part of the protocol we use perturbation theory in $(J_e/\Delta)$ to obtain the approximate evolution
\begin{align}
U_1' = e^{iJ_et/2} \begin{pmatrix}
e^{i \Delta t /2} & 0 \\
0 & e^{-i \Delta t /2}
\end{pmatrix}. \label{eq:udduevo}
\end{align}
On the other hand, the evolution in the $\{ \uuket, \ddket \}$ subspace  is given by
\begin{align}
U_2 = \begin{pmatrix}
e^{-i B_{tot} t/2} & 0 \\
0 & e^{i B_{tot} t/2} 
\end{pmatrix}, \label{eq:uuddevo}
\end{align}
where $B_{tot} = B_a + B_b$.  Now starting from the initial state $|\psi_0 \rangle = (\udket - \duket) | \! \uparrow\downarrow \cdots \rangle$ (suppressing the normalization of the state) and successively applying \swap s and the evolutions in Eq.~\eqref{eq:udduevo} and Eq.~\eqref{eq:uuddevo}, we find
\begin{align}
| \psi_0 \rangle \rightarrow | \psi_1 \rangle = i e^{i \Delta T_e /2} | \! \downarrow \uparrow \downarrow\uparrow \cdots \rangle - e^{i B_{tot} T_{e}/2} | \! \uparrow \downarrow \downarrow\uparrow \cdots \rangle
\end{align}

after the first period, where we used that $e^{iJ_eT_e/2}=i$, and we ignored accumulated phases coming from spins other than the first three.  The second period of the protocol yields
\begin{align}
| \psi_1 \rangle \rightarrow | \psi_2 \rangle = - ( \udket + \duket )| \! \uparrow\downarrow \cdots \rangle,
\end{align}
so that the first qubit is in the state $| T_0 \rangle$.  Two further periods then recover the initial state on sites 1 and 2, explaining the $4T$ periodicity of the singlet return probability.

\begin{figure}[h]
\includegraphics[scale=0.82]{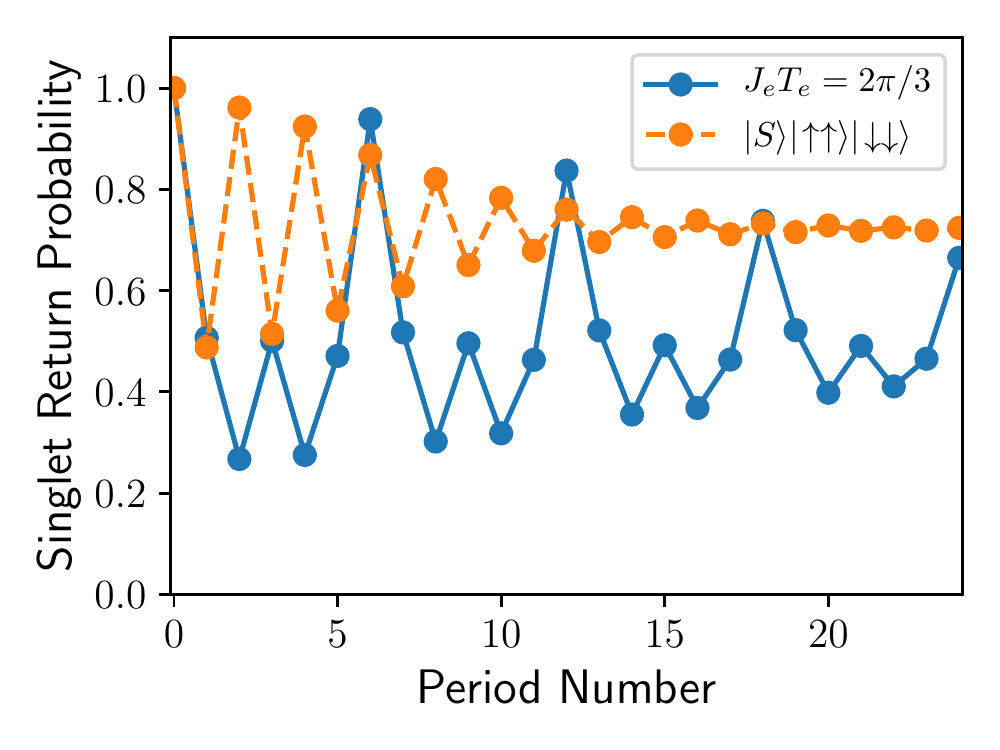}
\caption{ Singlet return probability for the cases in which the total phase accumulation of the evolution part of the protocol is $J_e T_e = 2 \pi/3$ (blue line) and in which the initial state is $|S\rangle|\! \uparrow \uparrow \rangle |\! \downarrow \downarrow \rangle$ (orange line).  In the first case, the initial state is the singlet state for qubit 1 and the product states minimizing the field gradient energies for the other qubits.  In the second case, $J_e T_e = \pi$.  Other parameters are $L=6$, $B_0 = 3075$ MHz, $\sigma_B = 18$ MHz, $T_e = 1.4$ $\mu$s, $T_S= 2$ ns, $J_S = \pi/T_S$, $\epsilon=0$.  Results are averaged over 6000 disorder realizations. \label{fig:retprobvst_extracases}}
\end{figure}

To provide further support for this simple physical picture, we consider two extensions of the idea.  We note the $4T$ periodicity fundamentally arises from the phase factor $e^{i J_e t/2}$ in Eq.~\eqref{eq:udduevo} becoming trivial after four periods, when $J_e T_e = \pi$ (here $J$ is given in radians and $\hbar = 1$).  Thus, one should obtain a different periodicity when $J_e T_e$ is chosen such that the relative phase winding occurs at another rate.  That this is indeed the case is shown in Fig.~\ref{fig:retprobvst_extracases}, where $J_e T_e = 2\pi/3$ and the resulting periodicity of the singlet return probability maxima is $6T$.  Alternatively, one may consider initializing the second qubit in the state $\uuket$ (with the first qubit still initialized in $|S\rangle$).  A similar argument as above shows that the first qubit returns to the singlet state after $2T$, in agreement with the orange curve in Fig.~\ref{fig:retprobvst_extracases}. In longer chains, a singlet state prepared in the bulk experiences $4T$ periodicity of the return probability at an interaction strength $J_e T_e = \pi/2$, half the value for a ST qubit on the edge.  This is essentially due to the increased number of neighbors, and mirrors the case of the single spin return probability, for which the phase diagram of a bulk spin has half the period compared to that for an edge spin \cite{Li2020}.

\section{Comparison with the Undriven System \label{sec:undriven}}

\begin{figure}[h]
\includegraphics[scale=0.48]{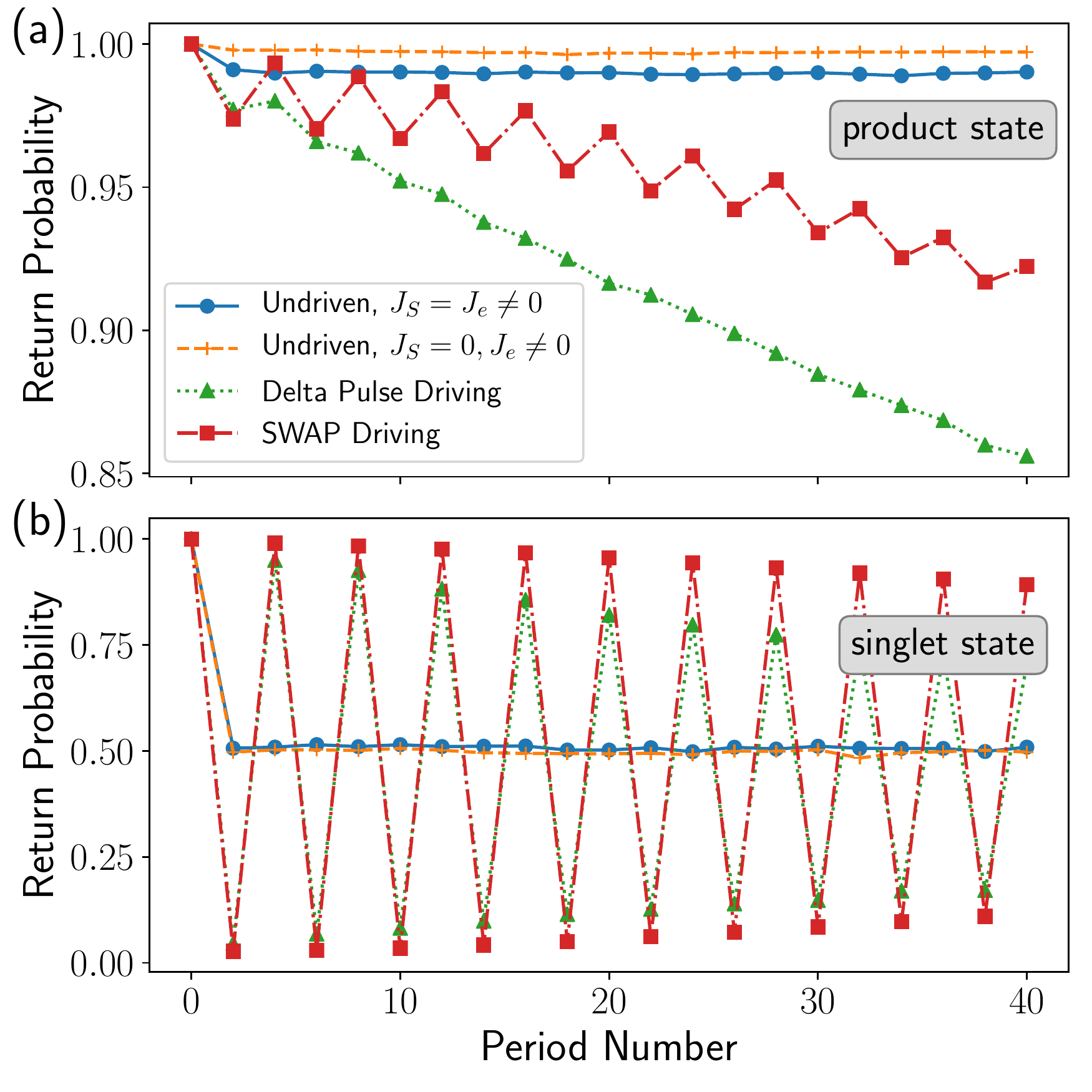}
\caption{ (a) Comparison between the undriven system and driving protocols, for an initial product state that minimizes the field gradient energy of qubit 1. (b) Comparison between the undriven system and driving protocols, for an initial singlet state of qubit 1. Other parameters are $L=4$, $B_0 = 3075$ MHz, $\sigma_B = 18$ MHz, $T_e = 1.4$ $\mu$s, $J_e = \pi/T_e$. For the \swap\ driving case $T_S = 2$ ns, $J_S = \pi/T_S$, and for both driven cases $\epsilon=0$.  The initial state of qubit 2 is the one minimizing the field gradient energy. Results are plotted stroboscopically for every $2T$ and averaged over 6000 disorder realizations. \label{fig:retprobvst_compareproductsinglet}}
\end{figure}

As noted in Section~\ref{sec:PDs}, the product state $\udket$ on qubit 1 is well-preserved even in the absence of \swap\ driving.  In Fig.~\ref{fig:retprobvst_compareproductsinglet}(a) we study the return probability as a function of time, for several different driving protocols.  Two different undriven cases are presented.  In the first, the Heisenberg interactions are equal throughout the chain and set to the same value as used for the \swap\ driving evolution: $J_{12}=J_{23}=J_{34}=\pi/T_{e}$.  However, since the \swap\ DTC evolution piece only involves inter-qubit $J_e$, the second undriven case mirrors this by setting $J_{23} = \pi/T_e$ and $J_{12} = J_{34} = 0$. In either case, while the undriven and \swap -driven cases perform similarly up to ten periods, in the long-time limit the undriven cases are clearly superior.   The saturation value of the return probability for the undriven cases tends to grow with increasing field noise strength \cite{Barnes2016}.  We note, however, that it does not ultimately approach 1 in the large noise limit.  This is due to the fact that disorder averaging mixes in unfavorable field configurations, which limits the overall return probability.  On the other hand, applying a uniform linear field gradient (not shown) does tend to increase the return probability towards 1, as the gradient strength increases.

We also compare the \swap\ protocol to more traditional single-spin driving.  Thus, we consider an idealized instantaneous $\pi$ rotation of all the spins (i.e. a delta-pulse in time):
\begin{align}
	V_\delta(t) = \frac{\pi}{2}   \sum_{s=1}^\infty \delta (t - s T ) \sum_{j=1}^L \sigma^x_j.
\end{align}
In this case, all nearest-neighbor exchange interactions are turned on, as in the first undriven case. The period of the delta-pulses is adjusted to coincide with the total period of \swap\ driving cases, $T_\delta = T_e + T_S$.  Fig.~\ref{fig:retprobvst_compareproductsinglet}(a) shows that for an initial product state, the \swap\ driving is preferable to the single-spin rotations of the delta-pulse case for experimentally relevant levels of magnetic field noise. 

Turning to the case where qubit 1 is initially in an entangled state, it is apparent from Fig.~\ref{fig:retprobvst_compareproductsinglet}(b) that an initial singlet state is not at all preserved for the undriven protocols, whereas the \swap\ case leads to a high return probability every four periods, in accordance with the results above.  In the given parameter regime, we again see that delta-pulse single-spin rotations are inferior to \swap\ pulses for preserving the initial state.

\begin{figure}[h]
\includegraphics[scale=0.4]{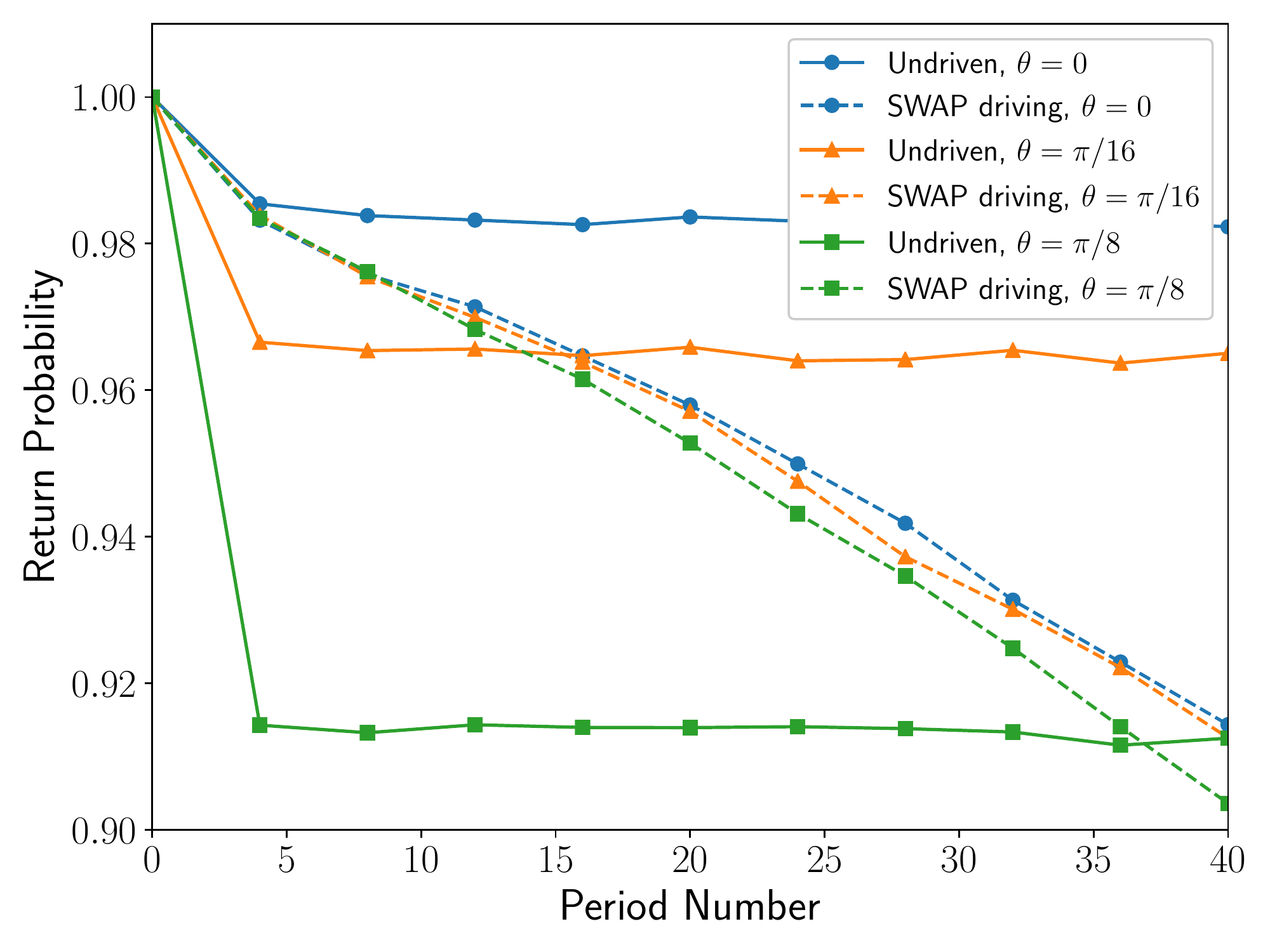}
\caption{ Comparison between the undriven (first case; all nearest neighbor interactions on) (solid lines) and \swap -driven (dashed lines) systems when qubit 1 is initialized in $\cos(\theta/2) \udket - \sin(\theta/2) \duket$. Results are shown for $\theta=0,\pi/16,\pi/8$.  Results are plotted stroboscopically every $4T$. Other parameters are $L=4$, $B_0 = 3075$ MHz, $\sigma_B = 18$ MHz, $T_e = 1.4$ $\mu$s, $J_e = \pi/T_e$. For the driven case, $T_S=2$ ns, $J_S = \pi/T_S$, $\epsilon=0$.  The initial state of qubit 2 is $\udket$.  Results are averaged over 6000 disorder realizations. \label{fig:retprobvst_unbalanced}}
\end{figure}

We have seen that the product states $\udket$ and $\duket$ survive longer in the absence of \swap\ driving, whereas $|S\rangle$ and $|T_0\rangle$ are preserved better when the system is driven. This suggests that if we consider ``unbalanced'' superpositions $\cos(\theta/2) \udket - \sin(\theta/2) \duket$ where $0<\theta<\pi/2$, there should exist some value $\theta_*$ such that for $\theta>\theta_*$, driving is beneficial for state preservation.  The value of $\theta_*$ in fact depends on how long one wishes to preserve the state, as is shown in Fig.~\ref{fig:retprobvst_unbalanced}.  The undriven system return probabilities depend strongly on $\theta$, but are essentially time-independent after an initial decay.  Here we have considered the first type of undriven system, in which all nearest-neighbor exchange interactions are nonzero and equal.  In contrast, \swap\ driving leads to a steady decay of the return probability as the number of driving periods is increased; this decay is relatively insensitive to $\theta$.  The intersection of the return probability curves for the undriven and \swap-driven cases yields the time below which \swap\ driving enhances the attainable return probability for a given initial state parameterized by $\theta$. Conversely, we may fix the time scale at a desired value and then read off the value of $\theta_*$ by adjusting $\theta$ until the undriven return probability curve intersects the \swap-driving curve at that time. Similar results are obtained for states with complex coefficients (not shown). Averaging over 88 states approximately distributed equally across the Bloch sphere, the undriven system yields a return probability of $0.65$ after 40 periods, compared to $0.90$ for the \swap\ driven case.  This indicates that a generic state is much better preserved by driving the system with the \swap\ DTC protocol.

\section{Switching preserved states \label{sec:switchstate}}

\begin{figure}[h]
\includegraphics[scale=0.5]{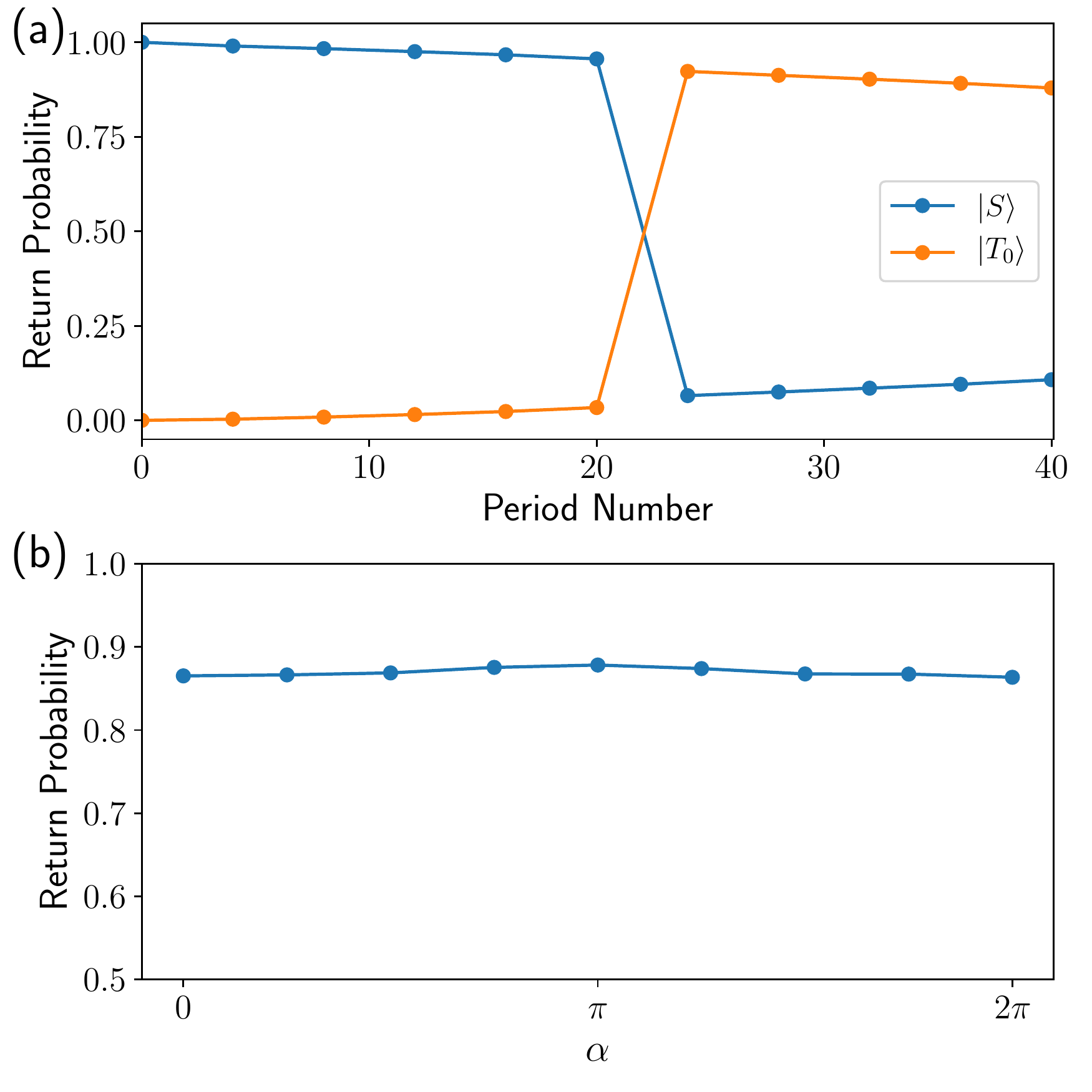}
\caption{ (a) Return probabilities for qubit 1 for the singlet (blue line) and triplet (orange line) states, for a system initialized in the singlet state for qubit 1 and subject to the switching protocol half way through the total evolution.  (b) End time return probability for the state $|\psi \rangle = \udket + e^{-i\alpha} \duket$ for qubit 1, when it is initialized in the singlet state and subject to the switching protocol half way through the evolution.  Other parameters are $L=4$, $B_0 = 3075$ MHz, $\sigma_B = 18$ MHz, $T_e = 1.4$ $\mu$s, $J_e = \pi/T_e$.  The initial state of qubit 2 is $\udket$.  Results are averaged over 6000 disorder realizations.\label{fig:switchstateandvaryalpha}}
\end{figure}

In the course of an information processing task, it is necessary to be able to change what state is stored in the memory.  In Fig.~\ref{fig:switchstateandvaryalpha}(a) we show that an initial $|S\rangle$ state, preserved for 20 periods, can be switched to the $|T_0\rangle$, and subsequently preserved to a similar degree.  The switching operation is performed simply by inserting an additional two periods with $J_e = 0$, halfway through the experimental run.  

More generally, one can switch from $|S\rangle$ to an arbitrary state of the form $\udket + e^{i \alpha} \duket$ by adjusting the value of $J_e$ during the two extra periods, such that $J_e T_e = \alpha$.  Fig.~\ref{fig:switchstateandvaryalpha}(b) shows that the return probability for the new state after $\sim 40$ total periods of evolution remains large, regardless of the choice of $\alpha$.

\section{Implementing Two-qubit Gates \label{sec:twoqubitgates}}

While the preservation of quantum states is an important task for quantum computing, it is also necessary to manipulate states and execute various logical gates.  Here we explore the possibility of using the \swap\ driving protocol to realize two-qubit gates in a chain of ST qubits.  We first note that when qubit 1 is initialized in a singlet state, the return probability oscillates with period $4T$ ($2T$) if qubit 2 is in state $\udket$ ($\uuket$). This implies that the evolution after two periods is equivalent (up to single-qubit rotations) to a \textsc{cnot} gate, where qubit 1 is the target, and qubit 2 is the control, since qubit 1 flips from $|S\rangle$ to $|T_0\rangle$ depending on whether the spins in qubit 2 are parallel or antiparallel. However, this approach suffers from the disadvantage that parallel spin states are not part of the computational subspace of ST qubits. Conditional control of individual spins using ESR or EDSR would alleviate this issue by allowing one to temporarily map $\duket \rightarrow \ddket$ to execute the \textsc{cnot}, before restoring the $\duket$ state of the control bit.

Another approach is based on the effective Ising Hamiltonian between exchange-coupled ST qubits in a linear array \cite{Wardrop2014}.  An Ising interaction of the appropriate duration can be converted to a \textsc{cz} gate by applying additional single-qubit rotations \cite{Jones2001}:
\begin{align}
    \textsc{cz} = e^{-i \pi/4}e^{i \pi \sigma^z_1/4}e^{i \pi \sigma^z_2/4}e^{-i \pi \sigma^z_1 \sigma^z_2/4}
\end{align}

This suggests viewing the protocol for the \swap\ time crystal not only as a means of state preservation, but also as a way to generate two-qubit gates.  Indeed, whereas two periods of the protocol $U$ of Eq.~\eqref{eq:swapDTCprotocol} yield the best state preservation when $J_eT_e = \pi$ (for product states of a single qubit), setting $J_eT_e = \pi/2$ produces a \textsc{cz} gate when followed by  single-qubit rotations on each ST qubit, due to the effective Ising interaction between the ST qubits.  Later, we compare this two-period gate to one that uses a single period of \swap\ DTC evolution. We numerically study the \cz\ protocol in the $L=4$ spin chain, configured as two ST qubits.   The accuracy of the proposed gate can be assessed by looking at the probability of finding the evolved spins in the state that would be obtained from an ideal \textsc{cz} gate: $p_{\textsc{cz}} = | \langle \textsc{cz}_{ideal,i} | \textsc{cz}_{actual,i} \rangle |^2$.  Here, $|  \textsc{cz}_{actual,i} \rangle = U_{\textsc{cz}} | \psi_i \rangle$ and $|  \textsc{cz}_{ideal,i} \rangle = \textsc{cz} | \psi_i \rangle$, where truncation of the state to the logical subspace is implicit. The physically implemented gate is given by
\begin{align}
U_{\textsc{cz}} = \mathcal{R}^{(1,2)}_{z,\pi/2} [U_{\swap}(T_S) U_{evo}(T_e)]^2, \label{eq:UCZ}
\end{align}
where the exchange coupling $J_e$ in $U_{evo}$ is such that $J_e T_e = \pi/2$, while $J_S$ in $U_{SWAP}(T_S)$ remains the value required for a \swap\ operation: $J_S T_S = \pi$. The operation $\mathcal{R}^{(1,2)}_{z,\pi/2}$ implements a simultaneous rotation on each qubit by $\pi/2$.

The fact that $U_{\textsc{cz}}$ approximates a $\textsc{cz}$ gate can be seen by noticing that in the physically relevant parameter regime where $J_S\gg\Delta$ and $J_e\ll\Delta$, where $\Delta$ is the magnetic field gradient across neighboring QDs, the evolution (truncated to the logical subspace) after two periods is approximately given by 
\begin{align}
    [U_{SWAP}(T_S) U_{evo}(T_e)]^2 \approx \begin{pmatrix} i & 0 & 0 & 0 \\
    0 & 1 & 0 & 0 \\
    0 & 0 & 1 & 0 \\
    0 & 0 & 0 & i
    \end{pmatrix}\label{eq:2periodEvol}
\end{align}
in the basis $\{|0\rangle |0\rangle,|0\rangle |1\rangle ,|1\rangle |0\rangle,|1\rangle |1\rangle\}$, with $|0\rangle = \udket$ and $|1\rangle = \duket$ forming the logical basis of the ST qubits. This result can be obtained using the approximate expressions for each piece of the evolution given in Sec. \ref{sec:retprob}. The subsequent application of the $z$ rotations on each ST qubit as indicated in Eq.~\eqref{eq:UCZ} converts the right-hand side of Eq.~\eqref{eq:2periodEvol} into a \textsc{cz} gate. Below, we show that the discrepancy between $U_{\textsc{cz}}$ and $\textsc{cz}$ is mostly due to additional single-qubit gates that arise from terms of order $\Delta/J_S$ and $J_e/\Delta$. Thus, $U_{\textsc{cz}}$ remains locally equivalent to a $\textsc{cz}$ gate even when these higher-order effects are included.

\begin{figure}[h]
\includegraphics[scale=0.6]{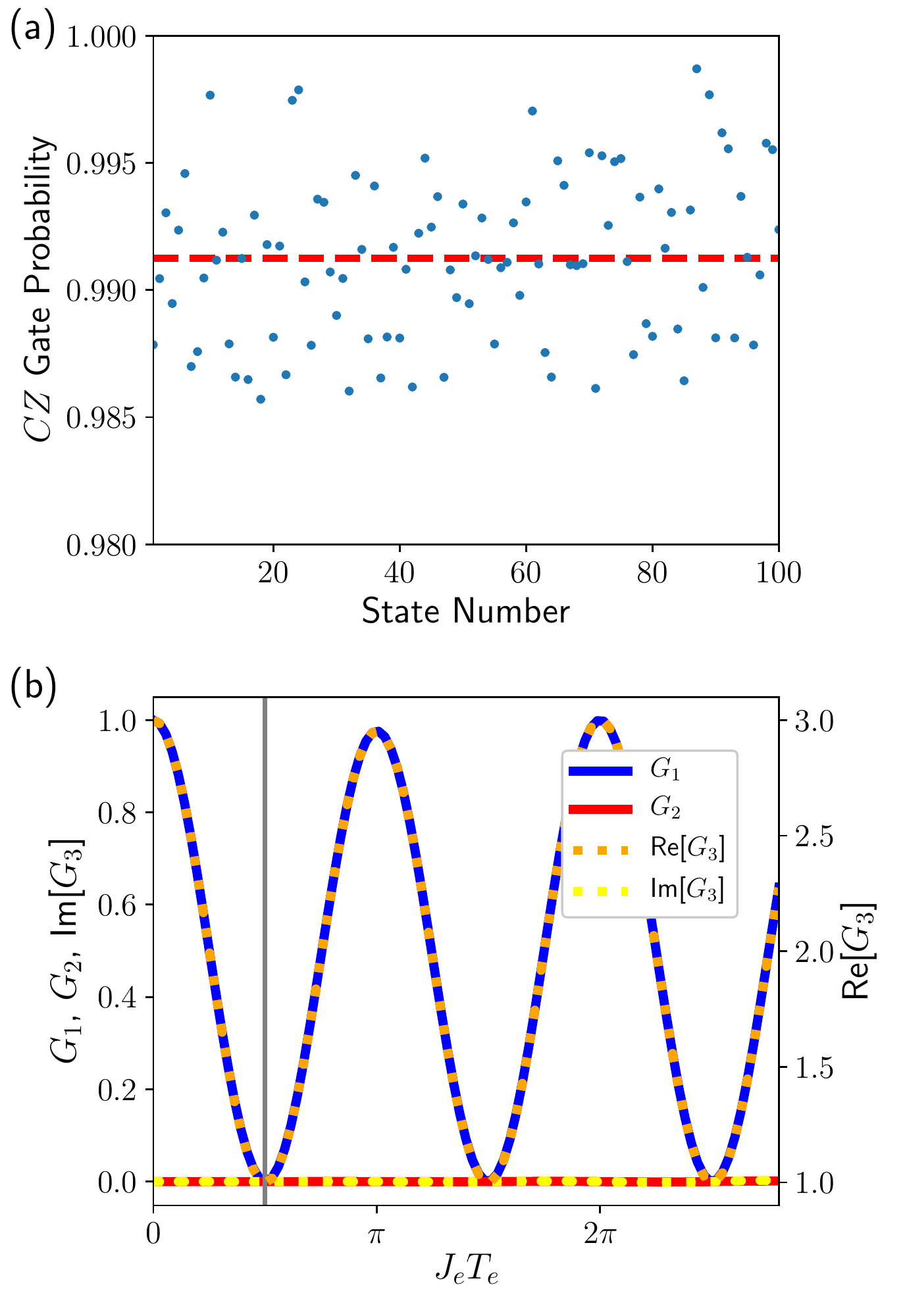}
\caption{(a) Probability $p_{\textsc{cz}} = | \langle \textsc{cz}_{ideal} | \textsc{cz}_{actual} \rangle |^2$ of finding the spin chain in the state that would be produced by  an ideal \textsc{cz} gate after the sequence in Eq.~\eqref{eq:UCZ} is applied. $J_e = \pi/(2T_e)$, with $T_e=1.4$ $\mu$s, as indicated by the vertical gray line in panel (b). Initial states are random product states in the ST qubit logical subspace.  The red dashed line indicates the mean CZ gate probability $\overline{p_{\textsc{cz}}}=0.991$. (b) Makhlin invariants $G_1$, $G_2$, and $G_3$, as functions of the inter-qubit coupling $J_e$.  Other parameters are $L=4$, $\Delta B_1=18$ MHz, $\Delta B_2=7$ MHz, $T_e=1.4$ $\mu$s, $T_S=2$ ns, $J_S = \pi/T_S$, $T_{g}=1$ $\mu$s. \label{fig:CZretprobmakhlin}}
\end{figure}

In Fig.~\ref{fig:CZretprobmakhlin}(a) we present numerical results for the $\textsc{cz}$ gate probability, $p_{\textsc{cz}}$, for 100 randomly selected initial product states of the ST qubits: $| \psi_i \rangle = | \psi^{(1)}_i \rangle | \psi^{(2)}_i \rangle$. Despite the single-qubit gates caused by finite $\Delta/J_S$ and $J_e/\Delta$, the mean probability is high: $\bar{p}_{CZ} = 0.991$. The use of more complicated pulse shaping techniques that effectively remove these extra local gates can be expected to improve this result further \cite{Wang2012,Barnes2015,Zeng2019a}. Unless noted otherwise, calculations are performed with fixed field gradients across each ST qubit, without any ``noise'' component. Corrective pulse shaping can be designed using the knowledge of these gradients to produce a pure $\textsc{cz}$ gate. In our simulations, the $\mathcal{R}^{(1,2)}_{z,\pi/2}$ operation is implemented by allowing each ST qubit to precess freely under its respective field gradients for a time $(T_{g} - t_r - 2T_S)/2$.  Here $T_{g}=1$ $\mu$s is the total gate time, while 
\begin{align}
t_r = \begin{cases}
\pi/(2\Delta) & \mathrm{if}\; \Delta > 0\\
3\pi/(2\Delta) & \mathrm{if}\; \Delta <0
\end{cases}
\end{align}
After this precession, a \swap\ pulse is applied and the qubit is allowed to precess again until $T_g - T_S$, at which time a final \swap\ is applied. This process allows for the rotation of the single-qubit state, along with an additional spin-echo-like part that keeps the different qubits in sync. Below, we also consider the noisy situation in which the true values of the gradients deviate from the ones assumed by the experimentalist implementing the gate. 

To assess the intrinsic entangling properties of the physical two-qubit \textsc{cz}, we compute the Makhlin invariants $G_1$, $G_2$, and $G_3$, which characterize a given two-qubit gate up to arbitrary single-qubit rotations \cite{Makhlin2002,Zhang2003}.  The Makhlin invariants for an ideal \cz\ are $G_1 = G_2 = 0$ and $G_3 = 1$.  Fig.~\ref{fig:CZretprobmakhlin}(b) shows the Makhlin invariants for the physical \cz\ as functions of the inter-qubit coupling $J_e$.  For the optimal value $J_e = \pi/(2T_e)$, the values of $G_{1,2,3}$ are given in Table~\ref{tab:makhlin}. One sees that the invariants of the physical gate closely approximate those of the ideal one.  This suggests that errors in the single-qubit rotations are the main factor leading to the imperfect \cz\ probabilities shown in Fig.~\ref{fig:CZretprobmakhlin}(a).  We also note that $G_3$ is necessarily real for any two-qubit gate.  Thus, the small imaginary part in the numerical calculation must arise due to leakage out of the computational subspace.  Fig.~\ref{fig:CZretprobmakhlin}(b) indicates that significant departures from the optimal $J_e$ lead to non-negligible errors in $G_1$ and $G_3$.  Thus, precise experimental control over the magnitude of $J_e$ is important for realizing the desired gate.  For a value of $J_e$ that is 1\% larger than optimal one, however, $G_3$ remains well within 0.01\% of its ideal value.
\begin{table}%[h!]
\begin{tabular}{|c | c | c | p{6em} |} 
 \hline
  & $G_1$ & $G_2$ & $G_3$ \\ [0.5ex] 
 \hline\hline
 Actual \cz\ & $3.5 \times 10^{-5}$ & $-4.1 \times 10^{-7}$ & 1 + $3.9 \times 10^{-5}$ $- 9.0 \times 10^{-7} i$ \\ 
 \hline
 Ideal \cz\ & 0 & 0 & 1 \\
 \hline
\end{tabular}
\caption{Makhlin invariants for the \swap-DTC two-qubit \cz\ gate.  Parameters are the same as in Fig.~\ref{fig:CZretprobmakhlin}.  }
\label{tab:makhlin}
\end{table}

\begin{figure}[h]
\includegraphics[scale=0.55]{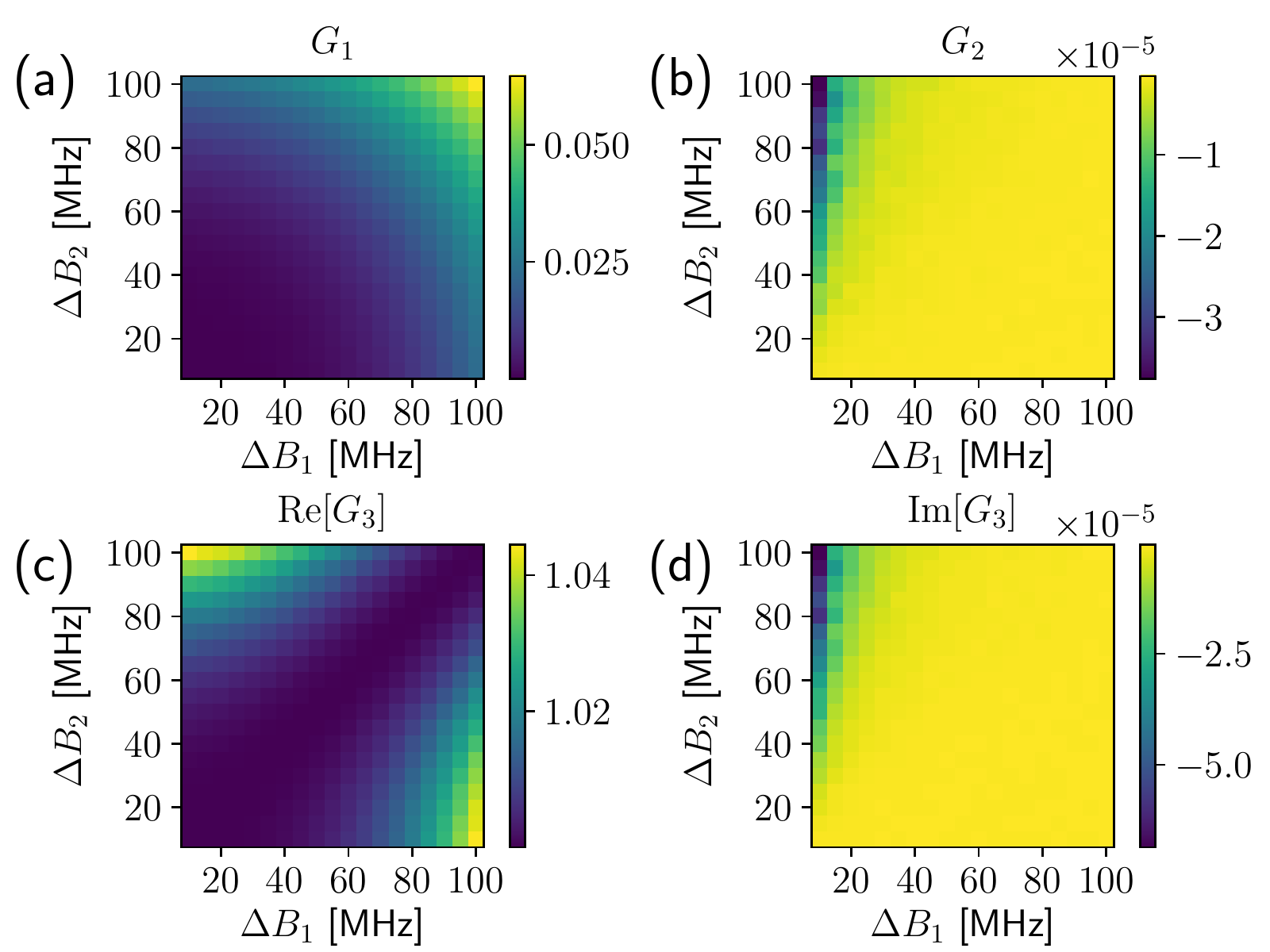}
\caption{ Makhlin invariants $G_1$, $G_2$, and $G_3$ as functions of the nominal magnetic field gradients across each qubit.  The true magnetic field for each data point is modified by the addition of Gaussian random field noise with standard deviation $\sigma_B = 1$ MHz.  Other parameters are $L=4$, $J_e=\pi/(2T_e)$, $T_e=1.4$ $\mu$s, $n_T=2$, $T_S=2$ ns, $J_S=\pi/T_S$,  $T_{g}=1$ $\mu$s.  Results are averaged over 40 disorder realizations. \label{fig:makhlinvsgrads}}
\end{figure}

One should also consider variations in the magnetic field gradients across the two qubits.  While these can be controlled to some extent, for instance, by micromagnet design, there are also contributions due to nuclear spin noise.  Fig.~\ref{fig:makhlinvsgrads} shows the Makhlin invariants for the physical \cz\ gate as functions of the magnetic field gradients across qubits 1 and 2, respectively (the left spins of each qubit are assumed to have the same field value).  In this figure, the axes give the nominal field gradients that are assumed in order to determine the pulse sequences that execute the necessary $z$ rotations.  The actual magnetic fields used in the calculation are modified, however, by the addition of Gaussian random field noise with standard deviation $\sigma_B = 1$ MHz.   The difference between the nominal and actual field values leads to errors in the single-qubit rotations of Eq.~\eqref{eq:UCZ}.  As the Makhlin invariants are unaffected by single-qubit rotations, the results are essentially the same as for $\sigma_B=0$ (not shown).  Nevertheless, we find that large values ($\sim 100$ MHz) of the field gradients lead to sizable departures from the ideal \cz\ gate, due to errors in the \swap\ gates induced by the gradients.  But for $\Delta B_1$, $\Delta B_2 < 50$ MHz, the Makhlin invariants remain close to the ideal ones.  Use of composite pulse shaping is expected to allow for successful operation in the larger gradient regime as well.

\begin{figure}[h]
\includegraphics[scale=0.5]{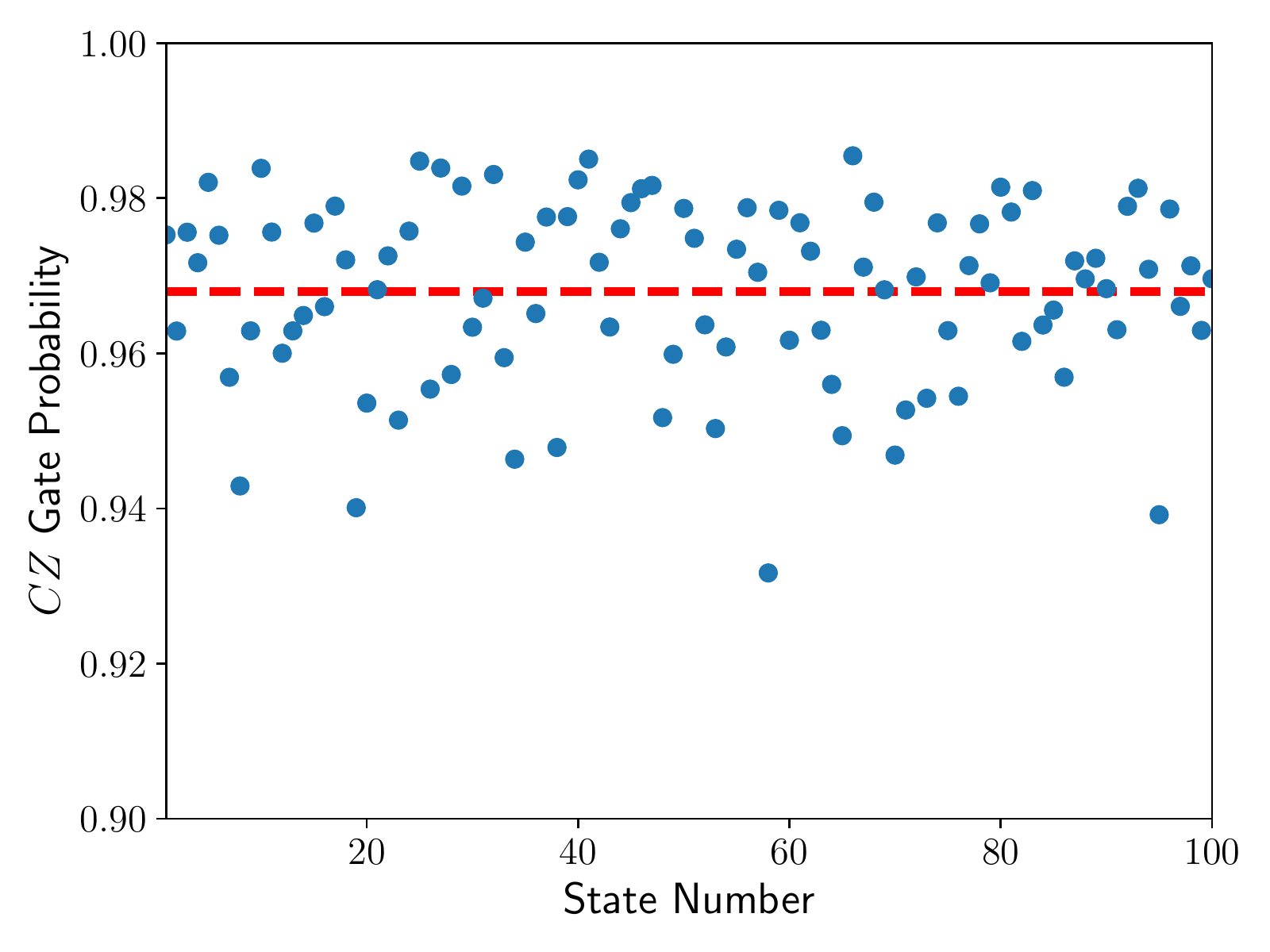}
\caption{\cz\ gate probabilities for random initial product states of the ST qubits, for which the true magnetic field for each data point is modified by the addition of Gaussian random field noise with standard deviation $\sigma_B = 1$ MHz. The red dashed line shows the mean value $\overline{p_{\textsc{cz}}}=0.968$. Other parameters are $L=4$, $\Delta B_1=18$ MHz, $\Delta B_2=7$ MHz, $J_e=\pi/(2T_e)$, $T_e=1.4$ $\mu$s, $n_T=2$, $T_S=2$ ns, $J_S = \pi/T_S$, $T_{g}=1$ $\mu$s.  Results are averaged over 20 disorder realizations. \label{fig:retprobBnoise}}
\end{figure}

Unlike the Makhlin invariants, the \cz\ gate probabilities are reduced by inaccurate $z$ rotations, and thus by differences between the nominal and actual magnetic field gradients in the system.  Fig.~\ref{fig:retprobBnoise} shows the return probabilities in the presence of $\sigma_B = 1$ MHz Gaussian field noise when the nominal gradients are $\Delta B_1 = 18$ MHz and $\Delta B_2 = 7$ MHz.  We find that the mean return probability is lowered from 0.991 in the noiseless case to 0.968 in the presence of noise.  This suggests that reliable knowledge of the field gradients is crucial for obtaining accurate ST qubit gates.

\begin{figure}[h]
\includegraphics[scale=0.55]{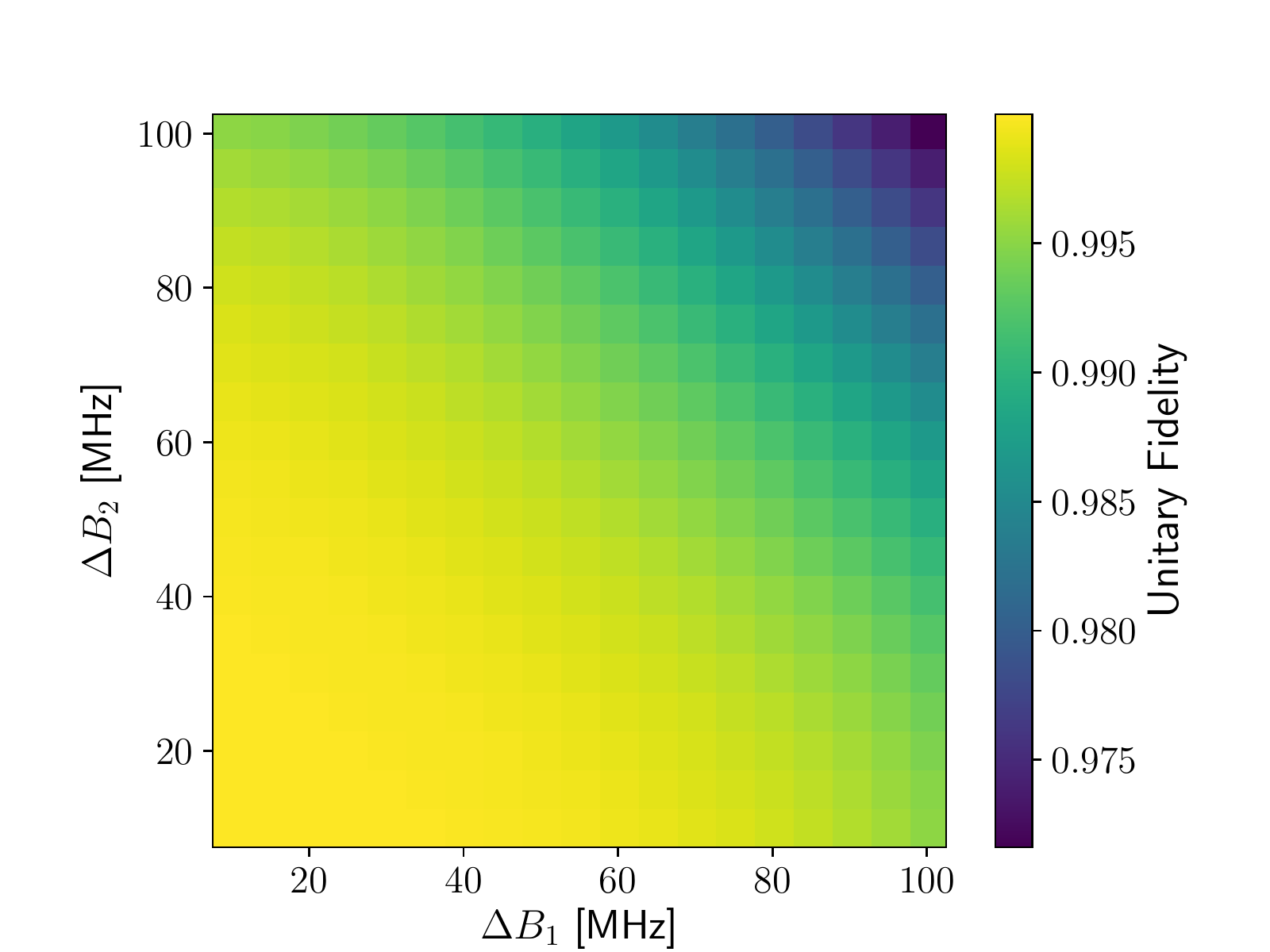}
\caption{ Optimized unitary fidelity of Eq.~\eqref{eq:unitaryfidelity} as a function of the magnetic field gradients across each qubit.  Other parameters are $L=4$, $J_e=\pi/(2T_e)$, $T_e=1.4$ $\mu$s, $n_T=2$, $T_S=2$ ns, $J_S = \pi/T_S$,  $T_{g}=1$ $\mu$s. \label{fig:optimizedfidelity}}
\end{figure}

An alternative metric for the quality of the physical \cz\ gate is given by the fidelity:  \cite{Pedersen2007,Economou2015}
\begin{align}
f = \frac{1}{20} (\Tr [U_{CZ,p} U_{CZ,p}^\dagger] + | \Tr [U_{CZ,p}^\dagger CZ^*]|^2), \label{eq:unitaryfidelity}
\end{align}
where $CZ^* = U_4 U_3 CZ U_2 U_1$ is the generalized \cz\ consisting of the ordinary \cz\ preceded by arbitrary one-qubit unitaries $U_{1,2}$ of the two qubits, and followed by the arbitrary unitaries $U_{3,4}$.  Furthermore, $U_{CZ,p}$ is the DTC part ($[U_{SWAP}(T_S) U_{evo}(T_e)]^2$) of the physical \cz\ gate projected down to the computational subspace, and $CZ^*$ is optimized over the parameters $\alpha_i, \beta_i, \gamma_i, \delta_i$ defining the one-qubit unitaries $U_i = e^{i \alpha_i} \mathcal{R}_z(\beta_i) \mathcal{R}_y (\gamma_i) \mathcal{R}_z (\delta_i)$.  With this definition, the optimized fidelity of the physical \cz\ gate is shown as a function of the magnetic field gradients in Fig.~\ref{fig:optimizedfidelity}.  For gradients below $50$ MHz, the optimized fidelity reaches values in excess of $0.995$, indicating that single-qubit rotations are the limiting factor in achieving an accurate gate in this case.  While $z$ rotations can be performed by turning off the intra-qubit exchange coupling $J_S$ for the appropriate length of time, thereby allowing the system to evolve in the ``always on'' field gradients, perfect $x$ rotations cannot be similarly achieved by applying a single value of $J_S$ for a given time, as the axis of rotation is tilted due to the gradients.  This again highlights the need for pulse shaping methods to improve single-qubit rotations.

\begin{figure}[h]
\includegraphics[scale=0.6]{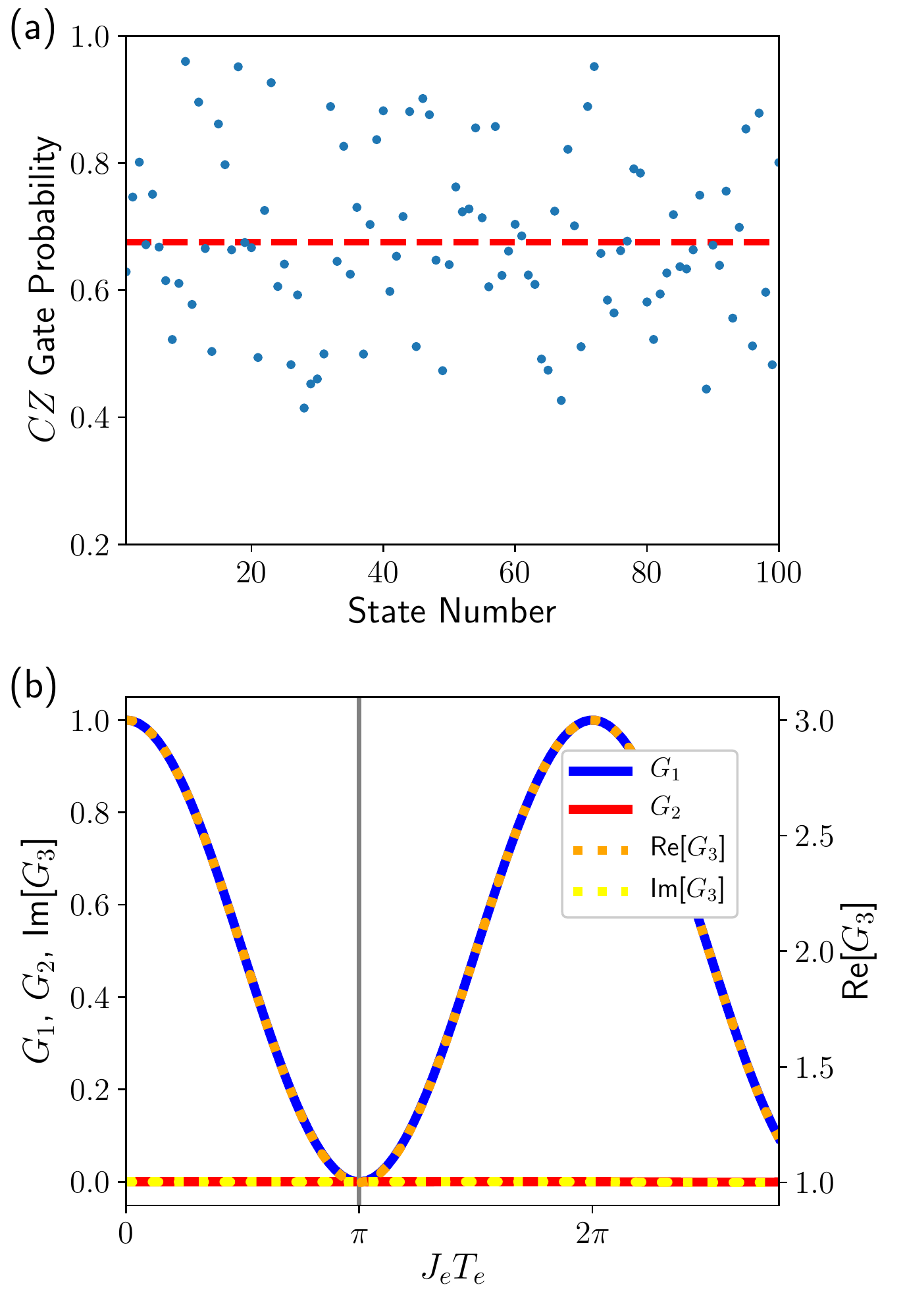}
\caption{ (a) \cz\ gate probability for the $n_T = 1$ protocol, using the optimal $J_e=\pi/T_e$, indicated by the vertical gray line in panel (b).  The red dashed line indicates the mean CZ gate probability $\overline{p_{\textsc{cz}}}=0.675$. (b)  Makhlin invariants for the $n_T=1$ protocol for the \cz\ gate. Other parameters are $L=4$, $\Delta B_1=18$ MHz, $\Delta B_2=7$ MHz, $\sigma_B=0$, $T_e=1.4$ $\mu$s, $T_{g}=1$ $\mu$s. \label{fig:CZretprobmakhlin_nT1}}
\end{figure}

\begin{figure}[h]
\includegraphics[scale=0.55]{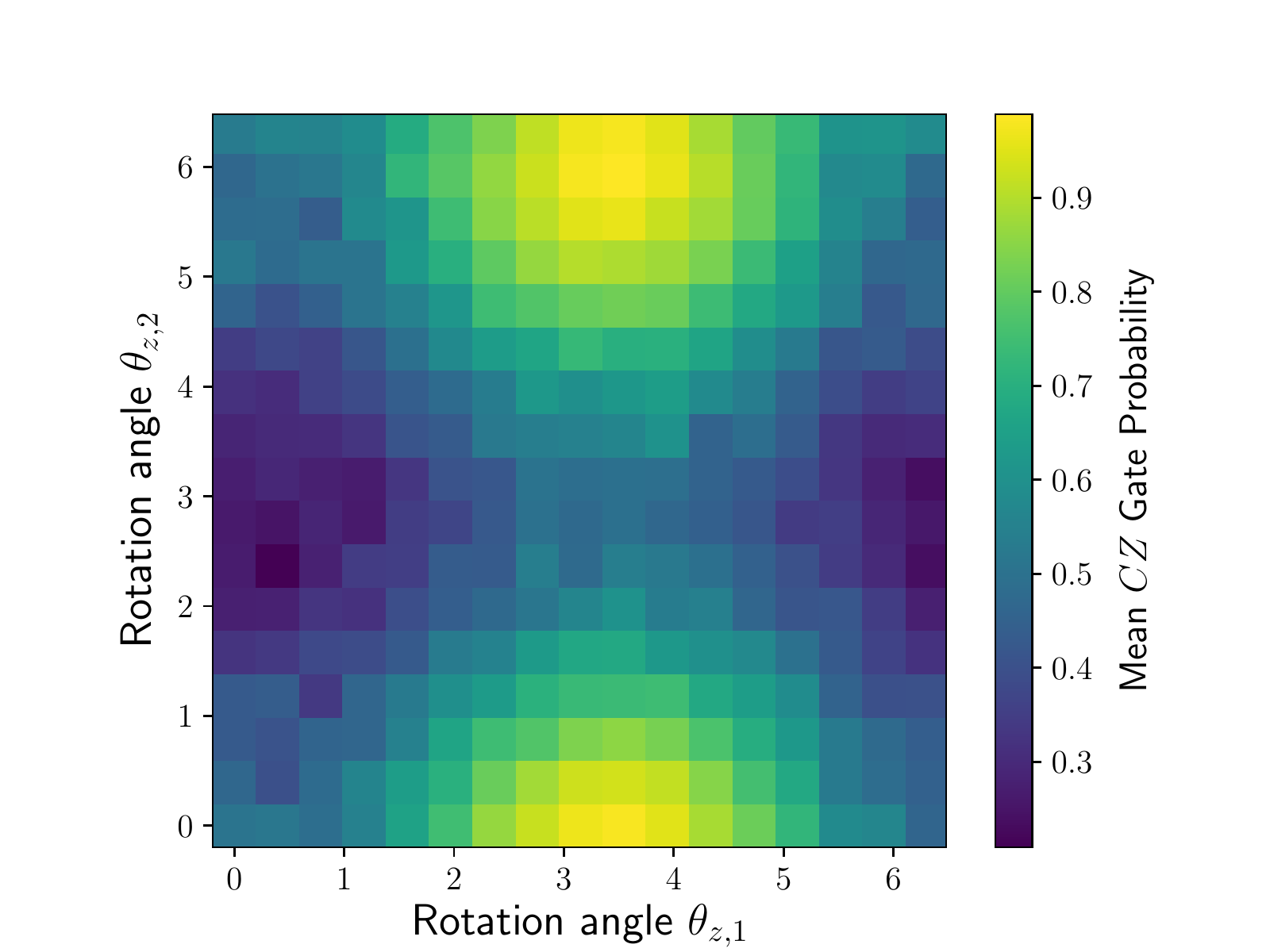}
\caption{ Mean \cz\ gate probability for the $n_T=1$ protocol, varying the single-qubit rotation angles applied after the two-qubit evolution. Other parameters are $L=4$, $\Delta B_1=18$ MHz, $\Delta B_2=7$ MHz, $\sigma_B=0$,  $J_e=\pi/T_e$, $T_e=1.4$ $\mu$s, $T_{g}=1$ $\mu$s.  Results are averaged over 100 randomly selected initial states, which are product states of generic ST qubit states.\label{fig:CZvaryangle_nT1}}
\end{figure}

Thus far we have considered a two-qubit \cz\ gate that requires two periods of the \swap\ DTC driving protocol, with a modified value of $J_e$ that maximizes the gate performance instead of preserving the initial state.  It is natural to ask whether a \cz\ gate could also be executed using a single period of inter-qubit evolution.  That is indeed the case, as illustrated in Fig.~\ref{fig:CZretprobmakhlin_nT1}(a), which shows that for a single evolution period such that $J_e T_e = \pi$, the Makhlin invariants are close to their ideal values.  Here, the evolution is not followed by the subsequent intra-qubit \swap\ pulses of the DTC protocol, as these amount to unnecessary additional single-qubit rotations.  However, the corresponding \cz\ gate probabilities for the optimal value of $J_e$ are very poor [Fig.~\ref{fig:CZretprobmakhlin_nT1}(a)].  This is due to the fact that the one-period protocol lacks the spin-echo behavior of the two-period version discussed above, which cancels the continuous $z$ rotations of ST qubits with finite field gradients.  Nevertheless, one can still achieve high \cz\ gate probabilities by selectively rotating each qubit through different angles $\theta_{z,1}$, $\theta_{z,2}$, such that the total rotation for each qubit at the end of the gate is the required $\mathcal{R}_{z,\pi/2}$.  This is seen in Fig.~\ref{fig:CZvaryangle_nT1}, which displays the \cz\ gate probability as a function of single-qubit rotation angles applied to each qubit after the inter-qubit evolution part of the gate.  The optimal choices of rotation angles depend on the field gradients across each qubit; in Fig.~\ref{fig:CZvaryangle_nT1} the highest return probability attained is 0.980, comparable to that of the two-period CZ protocol.

\begin{figure}[h]
\includegraphics[scale=0.55]{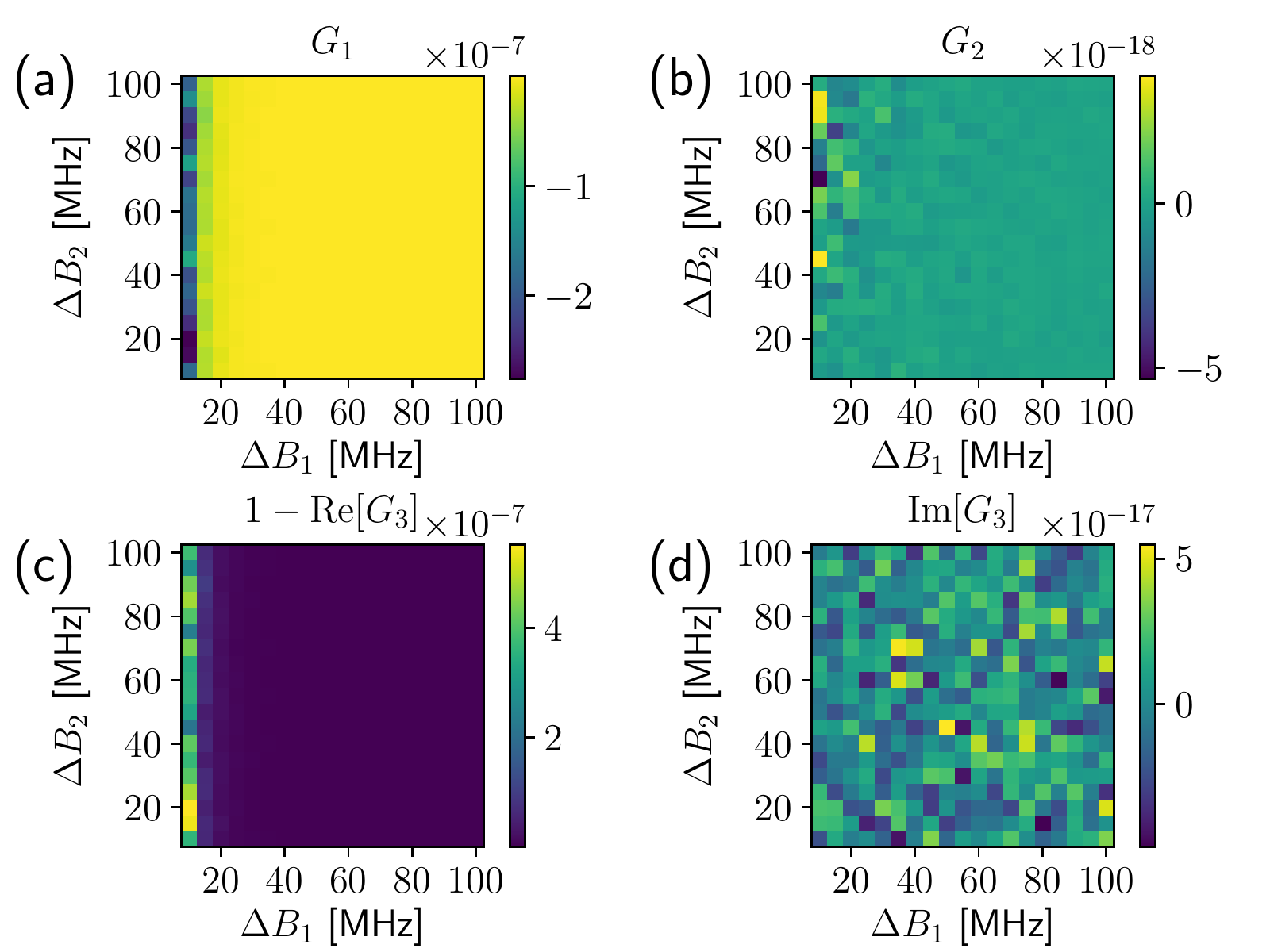}
\caption{ Makhlin invariants $G_1$, $G_2$, and $G_3$ as functions of the nominal magnetic field gradients across each qubit.  (Note that $1 - \mathrm{Re}[G_3]$ is plotted in (c)).  The true magnetic field for each data point is modified by the addition of Gaussian random field noise with standard deviation $\sigma_B = 1$ MHz.  Other parameters are $L=4$, $J_e=\pi/T_e$, $T_e=1.4$ $\mu$s, $n_T=1$, $T_{g}=1$ $\mu$s.  Results are averaged over 40 disorder realizations. \label{fig:makhlinvsgradsnT1}}
\end{figure}

\begin{figure}[h]
\includegraphics[scale=0.55]{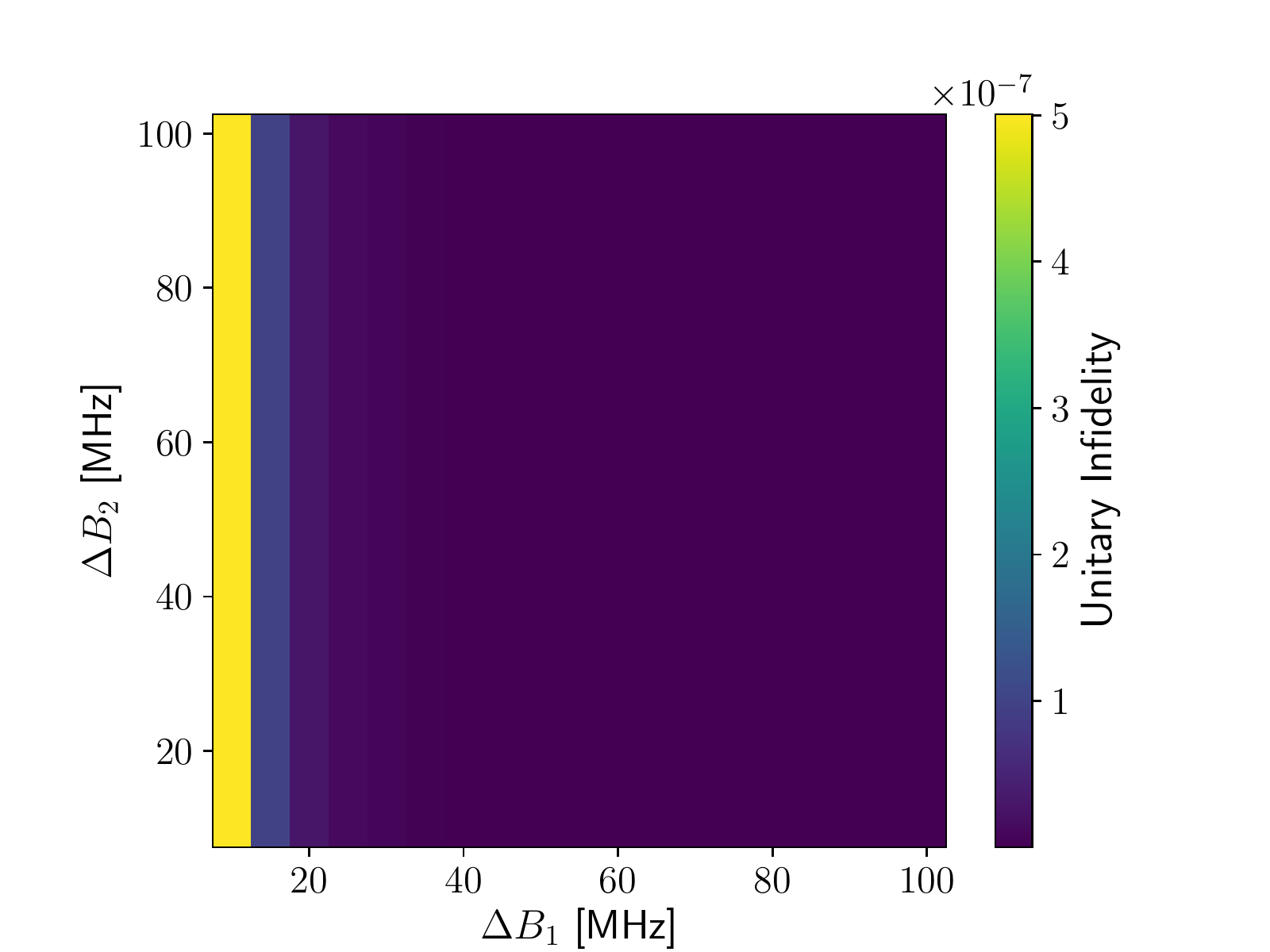}
\caption{ \cz\ gate infidelity for the one-period protocol ($n_T = 1$) as a function of the magnetic field gradients across each qubit. The infidelity at each point is optimized over single-qubit gate parameters. Other parameters are $L=4$, $J_e=\pi/T_e$, $T_e=1.4$ $\mu$s, $T_{g}=1$ $\mu$s. \label{fig:optimizedfidelitynT1}}
\end{figure}

The advantage of the one-period protocol (apart from the two-fold reduction in gate time) can be seen by considering the Makhlin invariants as functions of the magnetic field gradients [Fig.~\ref{fig:makhlinvsgradsnT1}].  The invariants remain within $10^{-5}$ of their ideal values throughout the range considered, thus showing considerable improvement from the two-period case at large gradients.  This suggests that optimizing over arbitrary single-qubit operations before and after an ideal \cz\ gate, in the manner of Eq.~\eqref{eq:unitaryfidelity}, should lead to very high fidelities. We confirm this expectation, as shown in Fig.~\ref{fig:optimizedfidelitynT1}, where the lowest infidelity over the range of gradients considered is only $\sim 5 \times 10^{-7}$.  Infidelities obtained in experiments will likely be higher due to single-qubit rotation errors.  Despite the significantly improved fidelities of the one-period protocol over the two-period version, the fact that the required $z$ rotations are gradient-dependent may present further experimental challenges.  This would necessitate adaptive control of the pulse sequence, in response to a prior measured value of the field gradient.  The two-period sequence, on the other hand, always involves $z$ rotations of $\pi/2$ for each qubit, regardless of the gradient strength, such that the pulse sequence does not need to be changed ``on the fly.''

\section{Conclusions \label{sec:conclusion}}

We have shown that driving exchange interactions, as opposed to performing single-spin rotations, in QD spin chains leads to an alternative route to time crystal physics that can be used for the preservation and manipulation of quantum states. We demonstrated that such driving is particularly useful for preserving the entangled singlet and triplet spin states often used as logical qubit states for quantum computation, and on average preserves arbitrary states on the Bloch sphere better than the undriven case. In addition, we uncovered additional signatures of the exchange-driven time crystal phase, including a $4T$ periodicity of the singlet return probability that runs counter to the $2T$ periodicity normally encountered in such systems. We also considered applications of this time crystal physics to the design of exchange-driven quantum gates for singlet-triplet qubits.  In particular, we showed that a simple modification of the \swap-DTC protocol yields a high-fidelity \cz\ gate, up to single-qubit operations.  These results suggest that time crystal physics may be beneficial to quantum information applications based on QD spin qubits.

\begin{acknowledgments}
We thank Bikun Li and Fernando Calderon-Vargas for helpful discussions.  This work is supported by DARPA Grant No. D18AC00025.
\end{acknowledgments}

\bibliography{bibprotectQIexchangedriving}

\end{document}